\documentclass[toc]{PoS}
\usepackage{fontenc,indentfirst,delarray,amsmath,amssymb}
\newtheorem{thm}{Theorem}[section]

\newtheorem{prop}[thm]{Proposition}

\newtheorem{defi}[thm]{Definition}
\newtheorem{rem}[thm]{Remark}

\newcommand{\norm}[1]{\left\lVert#1\right\rVert}
\newcommand{\abs}[1]{\lvert#1\rvert}

\newcommand{\scl}[2]{\langle#1,#2\rangle}
\newcommand{\suup}[1]{ \underset{#1}{\sup} }

\newcommand{\ket}[1]{\lvert#1\rangle}
\newcommand{\bra}[1]{\langle#1\lvert}

\def\lp{\left(} 
\def\rp{\right)} 
\def\dm{\lp\begin{array}}	
\def\fm{\end{array}\rp}
\def\m2{M_2 \lp \cc \rp}
\def\m3{M_3 \lp \cc \rp}	
	
\def\re{\text{Re}}			

\def\ot{\otimes}

\def\ds{\partial\!\!\!\slash}

\def\kk{{\mathbb{K}}}	
\def\cc{{\mathbb{C}}}
\def\C{{\mathbb{C}}}	

\def\R{{\mathbb{R}}}

\def\N{{\mathbb{N}}}
\def\zz{{\mathbb{Z}}} 		
\def\ii{{\mathbb{I}}}
\def\I{{\mathbb{I}}}

\def\ee{{\cal E}}
\def\E{{\cal E}}
	
\def\mm{{\mathcal M}}	
\def\M{{\mathcal M}}		
\def\aa{{\mathcal A}}

\def\A{{\mathcal A}}			
\def\bb{{\mathcal B}}			
\def\ll{{\mathcal L}}
\def\dd{{\mathcal D}}	
\def\hh{{\mathcal H}}

\def\pp{{\mathcal P}}
\def\ccc{{\mathcal C}}

\def\X{{\mathcal X}}

\def\ss{{\mathcal S}}

\def\BH{{\mathcal B}({\mathcal H})}

\def\ot{\otimes}

\def\xo0{\omega^0_x}
\def\yo0{\omega^0_y}

\def\akom{\alpha_\kappa\omega_m}
\def\akton{\alpha_{\tilde\kappa}\omega_n}

\def\xo0{x_\omega^0}
\def\yo0{y_\omega^0}

\def\pa{{\cal P}(\aa)}
\def\sa{{\cal S}(\aa)}

\def\fm{\Phi(x^\mu)}

\def\dm{\partial_\mu}

\def\X{{\cal X}}

\def\dmm{\left(\begin{array}}
\def\fmm{\end{array}\right)}

\newcommand{\HH}{\mathcal{H}}

\newcommand{\abso}[1]{|#1|}

\def\ds{\partial\!\!\!\!\!\slash}

\title{Length and distance on a quantum space}
\ShortTitle{Length and distance on a quantum space}

\author{Pierre Martinetti
\thanks{{Work supported by the {\bf ERC Advanced Grant} 227458 OACFT
  \emph{Operator Algebras \!\&\! Conformal Field Theory} and the
  {\bf ERG-Marie Curie fellowship} 237927 \emph{Noncommutative geometry \!\&\!
    quantum gravity}.}}
\\
        Dipartimento di Matematica e CMTP, Universit\`a di Roma Tor
        Vergata I-00133\\Dipartimento di Fisica, Universit\`a di Roma
        Sapienza  I-00185\\
        E-mail: \email{Pierre.Martinetti@roma1.infn.it}}

\author{Luca Tomassini\\
        Dipartimento di Scienze, Universit\`a di
Chieti-Pescara G. d'Annunzio  I-65127\\
        E-mail: \email{tomassini@sci.unich.it}}

\abstract{This contribution is an introduction to the metric aspect of
  noncommutative geometry, with emphasize on the Moyal
  plane. Starting by  questioning ``how to define a standard meter
  in a space whose coordinates no longer commute ?'', we list several
  recent results regarding Connes's \emph{spectral distance} calculated between eigenstates of the
  quantum harmonic oscillator \cite{Cagnache:2009oe}, as well as
  between coherent states \cite{Martinetti:2011fko}. We
  also question the difference (which remains hidden in the
  commutative case) between the spelral distance and the notion of \emph{quantum length}
  inherited from the length operator defined in various models of
  noncommutative space-time (DFR and $\theta$-Minkowski). We recall
  that a standard procedure in noncommutative geometry, consisting in
  doubling the spectral triple, allows to fruitfully confront the spectral
  distance with the quantum length. Finally we refine the idea of
  discrete vs. continuous geodesics in the Moyal plane, introduced in \cite{Martinetti:2011fk}.}

\FullConference{Proceedings of the Corfu Summer Institute 2011 School and Workshops on Elementary Particle Physics and Gravity\\
		September 4-18, 2011\\
		Corfu, Greece}

\begin{document}
\section{Introduction}
Till 1960 the standard meter was the distance  between two marks on
an iridium-platinum bar in  the \emph{Bureau des poids et mesures} in Paris. 
In 1960, the meter has been defined as $1\, 650\, 763.73$ wave lengths from an orange radiation
of $^{86}$Kr. Since 1983 \cite{SI2008} $1$ meter is the length of the path
travel\-led by light in vacuum during $\frac 1{299\, 792\, 458}$ of a second,
the latest being the duration of $9\, 192\, 631\, 770$ periods of
the radiation corresponding to the transition between the two hyperfine levels
 of the ground state of the caesium-$133$ atom. The definition of
 the meter thus relies on both special relativity - the speed of light is constant by
 definition - and quantum properties of nature - transitions between
 energy levels correspond to radiations with a fixed
 wavelength. Nowadays metrology mainly consists in
measuring the 
proper time interval{\footnote{All along the paper we use Einstein sum on repeated
    indices in alternate up/down position}}  
\begin{equation}
\Delta \tau  = \int d\tau = \frac 1c \int ds \quad\text{ with} \quad ds = \sqrt{g_{\mu\nu}(x) dx^\mu dx^\mu}  
\label{eq:8}
\end{equation}
between two suitably chosen events, like the emission and the reception of a light signal at one 
extremity of the object one is measuring (with a mirror reflecting
the signal at the other extremity).

The choice of the worldline along which
the integral (\ref{eq:8}) is performed becomes crucial as soon as the gravitational field is non-negligible with respect to the
uncertainties of the clock used to measure $\Delta \tau$. For instance, measuring by laser telemetry the
height of a tower of $400$m thanks to a clock localized at the
bottom or at the top of the tower yields two distinct values $h_{\text{top}} = \frac 12 c\Delta
\tau_{\text{top}}$,  $h_{\text{bottom}} = \frac 12 c\Delta
\tau_{\text{bottom}}$  which  differ in relative value by an amount bigger than the
uncertainty measurement of the best atomic clock
\cite{Guinot:1997fk}. Say differently, in presence of a sufficiently
non-uniform gravitational field, the length of an object is a frame
dependent notion.  To quote \cite{Guinot:1997fk}, the 1983
 definition of the meter ``fixes the unit of
proper length in the tangent three-dimensional space orthogonal to the
world line of the caesium atom providing the second (where the theory
of special relativity applies)'' and it is valid in  ''regions of space small enough that the non-uniformity of the
gravitational field has negligible effects with respect to the
uncertainties of measurement''. But if the precision of atomic-clocks were
improved by - say - two orders of magnitude, one could not
unambiguously measure the height of a $1$m tower. This would make the very notion
of ``meter'' problematic: two observers located at the
extremities $A, B$ of the (vertical)
iridium-platinum bar would disagree on the numerical value of the meter: is $1\text{m} = \frac 12 c \Delta\tau_A$ or $1\text{m} =\frac 12 c \Delta\tau_B$ ?  

In principle one could escape the problem by considering as a unit of
length a sufficiently
small submultiple of the meter, so that the non-uniformity of the
gravitational field becomes negligible. Not arbitrarily small however, since
at the Planck scale the classical picture of spacetime as
a smooth manifold is expected to loose
any operational meaning, due to the
impossibility of simultaneously measuring with arbitrary accuracy the
four spacetime coordinates $x_\mu$.  This comes  as a consequence of the
\emph{principle of gravitational stability against localization}
\cite{Doplicher:2001fk,Doplicher:2006uq}, which can be stated as follows:
\emph{The gravitational field generated by the concentration of energy
required by the Heisenberg Uncertainty Principle to localise an event
in spacetime should not be so strong to hide the event itself to any
distant observer - distant compared to the Planck scale.} 
In other term, to prevent the
formation of black-hole during an arbitrarily accurate localization
process, one postulates a non-zero minimal uncertainty in the \emph{simultaneous}
measurement of \emph{all} coordinates of space-time. 
Typically, in a flat four dimensional space-time $\M$ and assuming suitable symmetry for the measuring-probe, one
has \cite{Doplicher:1995hc}
\begin{equation}
  \label{eq:1}
  \Delta x_0 (\Delta x_1 + \Delta x_2 + \Delta x_3) \geq
  \lambda_P^2,\quad
\Delta x_1 \Delta x_2 + \Delta x_2 \Delta x_3 + \Delta x_3 \Delta x_1
\geq   \lambda_P^2
\end{equation}
where $\lambda_P$ is the Planck length. 
A way to
implement these uncertainty relations is to view the coordinates in a
chart $U$ of 
$\M$ no more as functions $x\in U\subset\M \mapsto x_\mu\in\R$,
but as quantum operators $q_\mu$ satisfying non trivial commutation relations,
\begin{equation}
  \label{eq:7}
  [q_\mu,q_\nu] = i\lambda_P^2 Q_{\mu\nu},
\end{equation}
where the $Q_{\mu\nu}$'s are operators whose properties 
depend on the model. 
Generalization to curve space-time have been investigated in \cite{Tomassini:2011fk,Doplicher:2012fk}.

Of course the accuracy of today's atomic-clocks is far too small
compare to the inhomogeneity of the gravitational field on Earth to make 
an effective measurement of $\lambda_P$ frame
dependant. However the 1983 definition of the unit of length relies in an
essential manner on
the differential nature of the line element \eqref{eq:8} and is
therefore incompatible with a quantum structure of spacetime as the
one postulated in  (\ref{eq:7}). To be more specific, assuming that
space-time at small scale is accurately described by quantum
coordinate operators $q_\mu$, how does one extract some metric information
from it ? None
of the objects entering formula (\ref{eq:8}) makes sense in a quantum
context. By this we mean:  

\vspace{0.25truecm}
\hspace{5.75truecm}
\begin{minipage}{.1\linewidth}
$\!\!\!?$\\ $\nwarrow$
\end{minipage}\hspace{-.8truecm}\begin{minipage}{.1\linewidth}
$\quad$\\
\vspace{-.5truecm}
$\!\!\rightarrow ?$
\end{minipage}
\vspace{-.5truecm}
$$
\hspace{2truecm}\int_x^y \quad\sqrt{g_{\mu\nu}(x) dx^\mu dx^\nu}
$$
\hspace{8truecm}\begin{minipage}{.1\linewidth}
\vspace{-.5truecm}$\swarrow$\\$\!\! ?$
\end{minipage}
\begin{minipage}{.1\linewidth}
\vspace{-.5truecm}
$\quad\;\searrow$\\
$\quad\quad\; ?$
\end{minipage}

\begin{enumerate}
\vspace{-.2truecm}\item [-] what would be a ``quantum metric tensor'' $g_{\mu\nu}$, and what
  would be its evaluation at some ``quantum point'' ?

\vspace{-.2truecm}\item [-] in case such an object is well defined (and there do exist
  proposals for a quantized version of the metric tensor, see e.g. \cite{madore}),
  how does one contract it with some ``quantum differentials'' $dq^\mu$
  so that to obtain a ``quantum line element''~?

\vspace{-.2truecm}\item [-] between which ``quantum points'' and along which ``quantum
  geodesics'' should one integrate this line element ?  And with
  respect to which theory of ``quantum integration'' ?
\end{enumerate}

In this contribution, we review two
proposals to answer these questions in the Riemannian
context. This does not allow to work out what could be the equivalent of the standard
meter in a quantum  \emph{space-time}, but it allows to give an
answer to all the question marks above for a quantum \emph{space}. These two proposals rely on an algebraic
definition of the length/distance, that makes sense in a noncommutative
framework and gives back the usual geodesic distance (i.e. the length
of the shortest path) when applied to the
commutative coordinates. The first proposal consists in defining a quantum \emph{length operator}
$L$, whose eigenvectors are interpreted as ``eigenstates of
length'' and the minimum $l_P$ of its spectrum is the
minimal value that may come out from a length measurement. As soon as
$l_P$ is non-zero, one inherits a natural notion of \emph{minimal length}. 
The second proposal is Connes' \emph{spectral distance} formula in noncommutative geometry.
\newline

Let us begin with the first proposal. As stressed above,  the quantum
coordinates satisfying (\ref{eq:7}) are well defined in the
flat case, so a natural candidate as a length operator is 
\begin{equation}
  \label{eq:1bis}
  L = \sqrt{\underset{\mu}{\sum} (dq_\mu)^2} \quad\text{ where } \quad dq_\mu \doteq q_\mu\otimes 1 - 1
  \otimes q_\mu,
\end{equation}
for it mimics the formula of the Euclidean distance (details in section \ref{sectioncom}). Explicit
computations of
\begin{equation}
l_P = \text{min} \left(\text{Sp}(L)\right)
\label{eq:40}
\end{equation}
have been made in 
\cite{Doplicher:1995hc,Bahns:2010fk} for the model  of
Doplicher, Fredenhagen and Roberts (DFR) in which the commutators
$Q_{\mu\nu}$'s are central operators with selfadjoint
closure, covariant under the action of the Poincaré group; as well as in \cite{Amelino-Camelia:2009fk} for the
canonical noncommutative space $\theta$-Minkowski where
the $Q_{\mu\nu}$'s  are constant. These results are recalled in
section \ref{sectionqlength}. 

Both models are
in fact viewed as quantum deformations of Minkowski spacetime, in that they
carry an action of either the usual Poincaré group (DFR), or a quantum group
deformation of it ($\theta$-Minkowski). However, regarding the computation of the \emph{quantum
length} (defined below as a mean value of $L$), these actions do not play any role (see
the appendix). Alternatively, one could wonder why we do not consider instead the Lorentzian operator
\begin{equation}
(dq_0)^2 -
\underset{\mu}{\sum} (dq_\mu)^2.
\label{eq:94}
\end{equation}
There are two difficulties: on the one hand the spectrum of this
operator is not bounded
below from zero and its physical interpretation as a quantum
observable is not transparent (see \cite{Bahns:2010fk});  on the other
hand the comparison with the
spectral distance formula would not be easy, since the latter makes sense only in the Riemannian case (although some
generalization to the Minkovskian case have been investigated in
\cite{Moretti:2003zw,Franco:2010fk}).
\newline

The second proposal is thus Connes spectral distance in
noncommutative geometry \cite{Connes:1989fk}. Before introducing it, let us
make a short digression:  in order 
to be interpreted as a
quantum-length ope\-rator with \emph{real} spectrum, the operator $L$
discussed above cannot be seen as an element of $\A_F\otimes \A_F$,
where $\A_F$ denotes the free algebra generated by relation \eqref{eq:7} and the identity $\I$. Indeed, as nicely
explained in \cite{Piacitelli:2010uq}, any element in $\A_F$ which is
not a multiple of $\I$ is a polynomial $p$ in the $q_\mu$'s, so that for
any complex number $\lambda$ one has
$(p-\lambda)^{-1}\notin \A_F$. 
This means that the spectrum of the element $p$ is the whole complex
plane, which makes the interpretation of the
$q_\mu$'s as physical observables difficult (a physical observable is
expected to have real spectrum). 
The point is thus to determine an Hilbert space $\HH$ on which the
$q_\mu$'s act as (unbounded) selfadjoint
operators. Their spectrum, 
\begin{equation}
\text{Sp}
(q_\mu) \doteq \left\{ \lambda\in\C, (q_\mu - \lambda\I) \text{ has no
  inverse in } \BH\right\}
\label{eq:92}
\end{equation}
would then be real, making the $q_\mu$'s acceptable physical
observables. Such an Hilbert space is obtained by viewing the $q_\mu$'s as operators
affiliated to a suitable noncommutative algebra $\A$. 
In the models of quantum spaces studied in this paper, $\A$ turns out to
be the algebra $\kk$ of compact operators (details
are given in section \ref{subsecalgebras}).

This digression illustrates how a suitably chosen noncommutative $*$-algebra
$\A$, viewed as the algebra of bounded continuous functions on the
quantum space, can be a more tractable (and chart independent) way to describe a quantum space than the algebra of
coordinates.  Such an idea is at the heart 
of Connes' approach to noncommutative
geometry \cite{Connes:1994kx} in which all the geometric
information is encoded within a \emph{spectral triple}, that is an operator $D$ - acting on some Hilbert space $\HH$ that carries a
representation of $\A$ - which generalizes the Dirac operator
$\ds = -i\gamma^\mu\partial_\mu$ of quantum field theory.
A distance on the state space $\sa$ of $\A$ (that is the set of positive, normalized, linear
applications from $\A$ to $\C$, see sections \ref{sectionpoints}) is defined by
\begin{equation}
  \label{eq:11}
  d_D(\varphi_1, \varphi_2 ) \doteq \underset{a\in\A}{\text{sup}} \{ \abso{\varphi_1(a)
  - \varphi_2(a)}, \norm{[D,a]}\leq 1\}.
\end{equation} This formula generalizes the Riemannian
distance to the noncommutative framework (proposition
\ref{propun}) and relies only on the spectral
properties of $\A$ and $D$. 
From a mathematical point of view, one can check without difficulty that (\ref{eq:11}) does define a
(possibly infinite) distance on $\sa$, that is a function
from $\sa\times \sa \to \R^+\cup\left\{\infty\right\}$ which is symmetric in the exchange of
its arguments, vanishes on the diagonal ($d_D(\varphi, \tilde\varphi)= 0$ iff
$\tilde\varphi = \varphi$) and satisfies the triangle inequality. 
\newline

In the following, we review several recent results on the spectral
distance and the length ope\-rator, mainly from
\cite{Cagnache:2009oe,Martinetti:2011fk,Martinetti:2011fko,DAndrea:2012fk}.
We begin by recalling in section 2 how to retrieve, in the commutative
case $Q_{\mu\nu}=0$, the
Euclidean distance from either the length operator $L$ or Connes'
formula \eqref{eq:11}. Then we go to the
noncommutative case and, under the assumption that the $Q_{\mu\nu}$'s are non-zero
 central operators, we single out in section 3 the Moyal algebra as a suitable algebra to describe the
quantum space (\ref{eq:7}).
In section 4 we make precise our notions of ``quantum points'' as pure
states of the $C^*$-closure of the Moyal algebra, which turns out to
be the algebra of compact operators
$\kk$. 
 In section 5 we discuss the quantum length $d_L$ of a two-''quantum
 point'' state $\tilde\omega\otimes\omega$, defined as 
 \begin{equation}
   \label{eq:55}
   d_L(\tilde\omega, \omega) \doteq (\tilde\omega\otimes \omega)(L).
 \end{equation}
We also recall the results on the spectral
distance $d_D$ for various classes
 of states of the Moyal algebra, including the eigenstates of the Hamiltonian of the
 quantum harmonic oscillator \cite{Cagnache:2009oe} and the coherent
 states \cite{Martinetti:2011fko}. 
Section 6 presents the strategy developed in \cite{Martinetti:2011fk} in
order to compare the quantum length
with the spectral distance, despite an obvious discrepancy (the latter
vanishes between a state and itself, the former does not as soon as
$l_P\neq 0$). This idea is to compare the quantum length with
the spectral distance on a double Moyal plane, that is the product of
the Moyal plane by $\C^2$. By a Pythagoras theorem for the product of
spectral triples, we show that this comparison is equivalent to
compare the spectral distance on a single Moyal plane with a new
quantity $d'_L$ build from the length operator, called the \emph{modified quantum length}.
The comparison of the spectral distance $d_D$ with the modified
quantum length $d'_L$ is the object of
 section 7. This is mainly the analysis developed
 in \cite{Martinetti:2011fk}, with further clarifications on the notion
 of discrete geodesics (between eigenstates of the harmonic oscillator)
 vs. continuous geodesics (between coherent states) in the Moyal plane.
\newpage

\section{Commutative case}
\label{sectioncom}

In this section, we recall how to retrieve the Euclidean distance from
the commutative coordinates $x_\mu$, either as the mean value of the
length operator $L$ on the state $\delta_x\otimes \delta_y$, or  as
the spectral distance $d_{\ds}(\delta_x, \delta_y)$. We begin with the algebraic characterization of the notion of
``points'', and close with a discussion on how the geodesic curves
emerge in this
picture.
  
\subsection{Points as characters}
\label{sectionpoints}
By Gelfand theorem, the
 points of a locally compact topological space $\X$ are retrieved as
 the characters of the commutative
 algebra $C_0(\X)$ of continuous functions vanishing at
 infinity. Recall that
a character is an algebra morphism between $C_0(\X)$ and
$\C$. Gelfand theorem simply means that rather than viewing a point $x\in\X$ as being
 acted upon by a function  $f\in C_0(\X)$  in order to give a number
 $f(x)$, a point can be equivalently viewed as the object that acts on a function $f$ to yields a
 number $\delta_x(f)$. This point of view is more compatible with
 quantum mechanics: classical physics assumes that space is the object
 that comes first, and functions acts on space to give number; in
 quantum mechanics there is no a priori given space, and observables come first.

Let us make some
mathematical comments, in order to prepare the generalization to the noncommutative setting
carried out in section ~\ref{subsecquantumpoints}. $C_0(\X)$ is a
$C^*$-algebra, whose elements have norm
\begin{equation}
\norm{f}\doteq \underset{x\in\X}{\sup}
\abs{f(x)}.\label{eq:9bis}
\end{equation}
 Characters are linear functionals on $C_0(\X)$ which are positive -
i.e. $\delta_x(\bar f f) = \bar f(x) f(x) \in\R^+$ - and normalized
- i.e. $\norm{\delta_x}=1$ - for the norm 
\begin{equation}
\norm{\delta_x} \doteq \sup_{f\in C_0(\X)}
\frac{\abs{\delta_x(f)}}{\norm{f}}.
     \end{equation}
Normalized positive linear functional on a $C^*$-algebra are called
\emph{states}. This is a
generalization of the notion of states in quantum
mechanics. Moreover, a general result of operator algebra gua\-rantees that the space of
state $\ss(\A)$ of any $C^*$-algebra $\A$ is convex, so that there
exist extremum elements, that is states that cannot be written as a
convex combination $\lambda \varphi_1 + (1-\lambda) \varphi_2$ of two
other states $\varphi_1, \varphi_2$. Such states are called
\emph{pure}. We denote $\pp(\A)$ the space of pure states of $\A$. In the commutative case,
characters are precisely the pure states of $C_0(\X)$:  Gelfand
theorem then reads
\begin{equation}
  \label{eq:35}
  \pp(C_0(\X))\simeq \X.
\end{equation}

\subsection{Quantum length and spectral distance}
\label{distoperator}

Let us consider the Euclidean space $\R^d$, $d\in\N$, with
 Cartesian coordinates $\{x_\mu\}_{\mu=1}^d$. 
Let $q_\mu$ denote the (unbounded, densely defined) selfadjoint coordinate operators whose action on $L^2(\R^d)$ reads
\begin{equation}
(q_\mu\psi) (x) \doteq x_\mu\psi(x).
\label{eq:26}
\end{equation}
The $q_\mu$'s do not belong to $C_0(\R^d)$ but are
affiliated{\footnote{\label{affiliated}An element $T$ is affiliated to a $C^*$-algebra $\A$ if
  bounded continuous functions of $T$ belong to the multiplier
  algebra $M(\A)$ of $\A$. In our context the unbounded operator
  $q_\mu$'s are affiliated to $C_0(\R^d)$, meaning that for any
  bounded continuous function $f$ on $\R^d$, $f(q_\mu)\in M(C_0(\R^d)) = C_b(\R^d)$ 
where $C_b(\R^d)$ is the algebra of bounded continuous functions on $\R^d$.}} to
it in the sense of Worono\-wicz \cite{Woronowicz:1991fk}.  The space being classical 
is traced back in the vanishing of the commutator $[q_\mu, q_\nu]$. 

\begin{prop}
\label{propun}
On pure states of $C_0(\R^d)$, the spectral distance \eqref{eq:11} associated to
 the spectral triple $(C_0^\infty(\R^d), L^2(\R^d), \ds)$ as well as the
quantum length $d_L$ introduced in \eqref{eq:55} coincide
 with the
Euclidean distance:
\begin{equation}
  \label{eq:152}
  d_{\ds}(\delta_x, \delta_y) = d_{\text{Eucl}}(x,y)  =
  d_L(\delta_x, \delta_y)\quad \forall x,y\in\R^d.
\end{equation}
\end{prop}

\noindent The proof is standard and can be found e.g. in
\cite{Martinetti:2011fk}. 

Notice that our definition of the length operator $L=\sqrt{\sum (dq_\mu)^2}$ heavily relies on the choice of the coordinate system: 
the $dq_\mu$'s are relevant only because the distance can be written as a
function of the difference of the coordinates, that is on a flat
space.

For the spectral distance, a more general result holds: viewing a
state{\footnote{Strictly speaking, one should talk of ``state'' only for
    $C^*$-algebras. In case $\A$ is not $C^*$, we consider states of
    its $C^*$-closure $\bar\A$ with respect to the operator norm
    coming from the representation on $\hh$. This is always possible,
    for from the axioms of spectral triples it follows that $\A$ is a
    pre-$C^*$ algebra.}} $\varphi$ of $C_0(\M)=\overline{C_0^\infty(\M)}$ as a probability measure $\mu$ on $\M$,
\begin{equation}
  \label{eq:2}
  \varphi(f) = \int_\M f d\mu,
\end{equation}
 then on a geodesically
complete Riemannian mani\-fold $\mm$ the spectral distance
$d_{\ds}$ between states coincides with the Wasserstein distance $W$ of
order $1$ between probabi\-lity distributions \cite{Rieffel:1999ec, dAndrea:2009xr}. The latter is defined in the
theory of optimal transport as the supremum of 
$\varphi_1(f) - \varphi_2(f)$ on all $1$-Lipschitz functions.  The
spectral distance being the same as the Wasserstein distance follows
from noticing that for any $f\in C_0^\infty(\M)$, then
\begin{equation}
\norm{[\ds,
  f]} = \underset{x\in\M}{\sup} \norm{\text{grad} \; f}\label{eq:5}
\end{equation}
is precisely the Lipschitz norm of $f$.
One then checks that any $1$-Lipschitz function on $\M$ can be approxi\-mated by a sequence of smooth
$1$-Lipschitz functions vanishing at infinity. For instance, for pure
states $\delta_x, \delta_y$, the supremum in the spectral distance
formula is attained by the function
\begin{equation}
  \label{eq:18}
  x\to d_{\text{geo}}(y, x),
\end{equation}
that we approximate by the
sequence 
\begin{equation}
  \label{eq:20}
 x\to \tilde d_{\text{geo}}(y, x)\, e^{-\frac{\tilde d_{\text{geo}}(y, x)}{n}}
\end{equation}
of functions in $C^\infty_0(\M)$, with $\tilde d_{\text{geo}}$ a smooth
approximation of $d_{\text{geo}}$. Notice that $\M$ being complete is
important to make  $\tilde d_{\text{geo}} $ vanish at infinity. 

\subsection{Geodesics}
\label{sectiongeodesic}

Although the length operator and the spectral distance are both
an ``algebraic version'' of the usual distance formula, the way of
addressing the problem is different:

- the definition (\ref{eq:1bis}) of the length
operator supposes that the Euclidean distance
\begin{equation}
  \label{eq:10}
  l(x_\mu) = \sqrt{\sum x_\mu^2}
\end{equation}
is known a-priori;

- the spectral distance formula (\ref{eq:11}) can
be seen as an equation whose solution
is (a suitable approximation of) the distance function $d_{\text{geo}}$.

More generally, given any spectral triple $(\A, \hh, D)$, we call  \emph{optimal element} between
two states $\tilde\varphi, \varphi$ an
element of $\A$ that attains the supremum in (\ref{eq:11}), or the
sequence of elements tending to this supremum in case the latter is
not attained. It is far from unique: given two points $x, y$ on a compact
manifold, any  function $d_{\text{geo}}(z,.)$ - with $z$ a point on the geodesic
between $x$ and $y$ not contained in the segment of curve between
$x$ and $y$ - is an optimal element between $\delta_x$ and $\delta_y$. 
On a non-compact complete manifold, an optimal element is
the sequence \eqref{eq:20}. We then call
$d_{\text{geo}}(z, .)$ an
optimal element \emph{up to regularization}.  

On the Euclidean plane, the function $l$ in
(\ref{eq:10}) is an optimal element up to regula\-rization between two points $x$
and $y = \lambda x, \,\lambda\in \R^+$. Identifying $C(\R^d)$
  with its representation on $L^2(\R^d)$, the optimal element up to regularization $l(q_\mu)$  and the  length
operator $L=l(dq_\mu)$ are the image of the same function $l$, under the functional
calculus of either the coordinates $q_\mu$, or their
universal differential $dq_\mu$. This formulation in terms of
functional calculus may sounds artificially  complicated in the
commutative case, but it is helpful to
understand the diffe\-rence bet\-ween the quantum length and the spectral
distance in the Moyal plane, where precisely the picture is different:
the function $l_1$ yielding the length operator no longer yields the
optimal element  (see section \ref{moyalgeo}). 

Notice also that, while a unique length operator $L$ gives the distance
between any $x,y\in\R^d$ as $(\delta_x\otimes\delta_y)(L) = \abs{x-y}$,
there is no function $\tilde l$ that would
be an optimal element between any two points: $\abs{\tilde l(x)
- \tilde l(y)}$ cannot equal $\abs{x-y}$ $\forall x,y\in\R^d$.
Furthermore, if $y\neq \lambda x$ then none of the optimal element
\begin{equation}
l_z(x_\mu)
\doteq \sqrt{\sum (x_\mu - z_\mu)^2}\label{eq:59}
\end{equation}
where $z$ is a fixed point such that $y-z =
\lambda (x-z)$  yields a length
operator, for
\begin{equation}
(\delta_x\otimes\delta_y) (l_z(dq_\mu)) = \sqrt{\sum (x_\mu - y_\mu -z_\mu)^2}\neq \abs{x-y}.
\end{equation}

\section{The algebras of quantum spacetime}
\label{sectiongeo}
A ``commutative'' space $\M$ can be characterized algebraically as the space of
pure states of the commutative algebra $C_0(\M)$. Similarly, a quantum
space may be defined as the set of pure states of a noncommutative
algebra $\A$. To find out the correct algebra associated with
the non-commuting coordinates operators $q_\mu$'s (\ref{eq:7}), one needs
to specify the property of their commutators $Q_{\mu\nu}$. In the
following, we make the assumption used in both the DFR model and
$\theta$-Minkowski, namely the $Q_{\mu\nu}$ are central:
\begin{equation}
[Q_{\mu\nu}, q_\alpha] =0 \quad\quad \forall \alpha,\mu,\nu.\label{eq:27}
\end{equation}
Doing so we exclude another model that
gained interest in the recent time: $\kappa$-Minkowski. 
As a matter of fact, the metric aspect of $\kappa$-Minkowski space has been
little studied, and is still an open problem, that we shall not
address here.
 
\label{subsecalgebras}

\subsection{The Moyal algebra}
\label{sectionschro}
A pathway to determine a ``natural'' $\BH$ on which the coordinate
operators $q_\mu$'s (\ref{eq:7}) act as unbounded operators comes from group
theory. Assuming the $Q_{\mu\nu}$'s are central and the
representation of the $q_\mu$'s is faithful and irreducible, by Schur lemma we write
\begin{equation}
  \label{eq:25}
  Q_{\mu\nu} = \theta_{\mu\nu}\ii,
\end{equation}
where $\Theta\doteq \left\{\theta_{\mu\nu}\right\}$ is an antisymmetric
matrix. We assume that $\Theta$ is non-degenerate, which forces the
dimension $d=2N$ to be even, and for $x,y\in\R^{2N}$ we denote 
\begin{equation}
\sigma(x, y) \doteq x^\mu\theta_{\mu\nu} y^\nu
\label{eq:29}
\end{equation}
the symplectic form induced by $\Theta$. Equation (\ref{eq:7}) then
defines the Heisenberg Lie algebra of dimension $2N$ with central
element{\footnote{Recall that given a symplectic vector
space $(V,\sigma)$ of real dimension $n$, the Heisenberg algebra with
central element $c$ is the
real central extension of the Lie algebra ${\mathfrak r}^n = \R^n$ of the additive group
$\R^n$, characterized by the relations
$[v,c] = 0, [v,v'] = \sigma(v,v')c$ for all $v,v'\in\R^n$.
 This is the Lie algebra of the Heisenberg group $H(V)$, namely the
 central extension $\R \ltimes V$ with group law
 similar to (\ref{eq:39}).  
}}
\begin{equation}
c= i\lambda_P^2.\label{eq:47}
\end{equation} By
exponentiation,
one gets the 
Heisenberg group
$H\doteq   \R^{2N} \ltimes \R\label{eq:3quat}$
with group law
\begin{equation}
  \label{eq:39}
  (x, \lambda).(x', \lambda') =
  (x+ x', \lambda +\lambda' +
  \frac 12 \sigma(x,x'))\quad \forall
  x, x'\in\R^{2N},\, \lambda, \lambda'\in\R.
\end{equation}

Now, to any locally compact group $G$ one naturally associates the enveloping
    $C^*$-algebra  $C^*(G)$, obtained by completing the Banach
    *-algebra $L^1(G)$
{\footnote{i.e. the completion with respect to the  the $L^1$-norm $\norm{f}_1 \doteq \int_G
\abso{f(t)} dt$ - $dt$ a Haar measure - of the algebra of compactly supported
    function on $G$, equipped with the convolution product
\begin{equation}
  \label{eq:23}
  (f\star g)(t)  \doteq \int_G f(s) g(t^{-1}s) ds 
\end{equation}
and  involution
$f^*(g) = \overline{f(g^{-1})}$.
}}
with respect to the norm 
\begin{equation}
\norm{f}_* \doteq \sup_\pi \{\norm{\pi(f)}\},
\label{eq:24}
\end{equation} 
where the supremum runs over
all representations of $L^1(G)$. Therefore it is tempting to consider $C^*(H)$ as the natural $C^*$-algebra associated to
the quantum space.
However a careful examination of the representations of
$L^1(H)$ indicates that $C^*(H)$ is too big:

- First of all in
\eqref{eq:24} it
is reasonable to take into account irreducible
representations only, since the $q_\mu$'s are known as soon as one has
determined their
irreducible action on $\HH$. 

- Then, recall that the
(irreducible) non-degenerate representations
of $L^1(H)$ are in $1$-to-$1$ correspondence with the (irreducible)  unitary
representations $(\pi, \HH_\pi)$ of $H$: to  any $f\in L^1(H)$
corresponds the bounded operator $\pi(f)$ defined by the Bochner
integral \cite{Deitmar:2009fk}
\begin{equation}
  \label{eq:34}
\pi(f)\doteq \int_{H_N} f(x,\lambda) \pi(x,\lambda) \, dx\,d\lambda
\end{equation}
acting on $\HH_\pi$ and any bounded non degenerate representation of
$L^1(H)$ comes in this 
way \cite{Deitmar:2009fk}.

- Irreducible unitary representations of $H$
are of two kinds, depending
on their central character $\chi${\footnote{i.e. the
homomorphism from the center  $Z(H)= (0,\lambda)
$ of $H$ to $S^1$ defined by $\chi(a)\I\doteq \pi(a)$.}}. Since $Z(H)\simeq \R$, any $\chi$
is of the form $\lambda~\mapsto~e^{i t\lambda}$ for a fixed $t\in \R$.  For $t=0$, the constant function $\chi(\lambda)~\doteq 1$ is the central character of an infinite number of nonequivalent
$1$-dimensional representations (see \cite{Deitmar:2009fk} for details).
For $t\neq 0$ one gets the
central character of the irreducible unitary representation on
$L^2(\R^{N})$ (unique up to equivalence, by von Neumann uniqueness theorem):
\begin{equation}
  \label{eq:32}
  \pi(x, \lambda) \varphi(u) \doteq e^{i t(x_2 u
    +\lambda)}\varphi(u + x_1) \quad\quad x= (x_1,  x_2)\in \R^{2N}, \lambda\in\R.
\end{equation}
Since $\pi(\text{exp} \, c) = \pi_t(0, 1) = e^{it}\I$ the value
of $t$ is fixed by (\ref{eq:47}) as $\doteq\lambda_P^2$.

Therefore, rather than \eqref{eq:24}  it is legitimate to consider the closure
of $L^1(H)$ with respect to the single representation with central
character $\lambda_P^2$. Let us denote this representation $\pi_\lambda$, with kernel $J$. 
As a Banach algebra, one gets  
\begin{equation}
  \label{eq:43}
  L^1(H)\slash J \simeq L^1(\R^{2N}\! ,\!\times)
\end{equation}
where $\times$  is the \emph{twisted convolution}
\begin{equation}
  \label{eq:28}
  (f \times g)(x) \doteq   \int_{\R^{2N}} f (x') g (x-x')\,
  e^{-\frac{i\lambda_P^2\sigma(x',x)}2} \, dz'  \quad\quad \forall f,g\in L^1(\R^{2N}).
\end{equation}
The 
norm closure of $\pi (L^1(\R^{2N})\slash J)$ is
$C^*(L^1(\R^{2N},\times))$, which turns out to be isomorphic to the
algebra of compact operators \cite{Doplicher:1995hc}
\begin{equation}
  \label{eq:50}
  C^*( L^1(\R^{2N}\!,\!\times)) \simeq \kk.
\end{equation}
 
To retrieve the algebra of compact operators from the traditional
Moyal product $\star$, let us recall that the latter is obtained as the pull-back through the
Fourier transform $F$ of the twisted
convolution (\ref{eq:28}),
\begin{equation}
  \label{eq:44}
f\star g  \doteq F^{-1}\left[F[f]\times F[g]\right].
\end{equation}
To close an algebra, one may for instance restricts to Schwartz
functions $f, g\in S(\R^{2N})$, for the twisted convolution, as the Fourier transform, maps
Schwartz function into Schwartz functions. 
Writing 
\begin{equation}
\theta \doteq \lambda_P^2,
\label{eq:69bis}
\end{equation}
standard Fourier theory 
yields the usual  form of the Moyal product, that is
\begin{equation}
  \label{eq:25bis}
  (f\star g)(x)  =
  \left(\frac{1}{\pi\theta}\right)^{2N}\int_{\R^{2N}\times \R^{2N}}du\,
  dv\, f(x+u) g(x+v) e^{{-\frac{2i}{\theta}}\, u\, \Theta_0^{-1}v},
\end{equation} 
where
\begin{equation}
\Theta_0=\left(\begin{array}{cc} 0 & \ii_N\\ -\ii_N &0 \end{array}\right).
\label{eq:63}
\end{equation}
\newline

Consequently, in the same way that $C_0(\M)$ is the natural $C^*$-algebra
associated to a manifold $\M$, the $C^*$-closure $\kk$ of the Moyal algebra
$\left(S(\R^{2N}),\star\right)$ {\footnote{The $C^*$-completion of the Moyal algebra in the
    operator norm coming from the (non-irreducible) left regular representation
    introduced below is a multiple of the $N$-fold ($C^*$) tensor product of $\kk$
    with itself, which is isomorphic to $\kk$. } }
is the natural algebra associated to the
quantum space (\ref{eq:7}) with central commutators.  As well, in the same way that the commutative coordinates $x_\mu$ do not belong
to $C_0(\R^d)$, the noncommutative coordinate operators $q_\mu$ do not belong to $\kk$, but they are
affiliated to it (see footnote p.7). This means that to find out the
desired representation of the $q_\mu$'s on the Hilbert space $\hh$ in equation
(\ref{eq:92}), one needs to study the representations of $\kk$. This is
the object of the next paragraph.

Notice that the analysis developed in this section does not take into
account the action of the Poincaré group (DFR model) or a deformed
version of it ($\theta$-Minkowski)
on the quantum coordinates. We explain in appendix why this action
does not play any role regarding the computation of distance and length.
\newline

\subsection{The left regular and the Schr\"odinger representations}
\label{subsectionschro}
The natural \emph{left regular} action $\ll$ of the
Moyal algebra  $(S(\R^{2N}),\star)$ on $L^2(\R^{2N})$,
\begin{equation}
  \label{eq:42}
  \ll(f)\psi = f\star \psi \quad\quad \forall \psi\in L^2(\R^{2N}),
\end{equation}
 is not irreducible (see e.g. \cite{Cagnache:2009oe}). As recalled in
 the next section, to characterize the pure states of the Moyal
 algebra, it is convenient to have an irreducible representation. 
The latter is nothing but the usual Schr\"odinger representation
$\pi_S$ of quantum mechanics. Let us work it out explicitly,
restricting ourselves to the Moyal plane
$N=1$ in order to fix notations.

On $\hh_S \doteq L^2(\R)$, we denote the position and momentum operators, 
\begin{equation}
\frak q: (\frak q\psi)(x) = x\psi(x),\quad \frak p:
(\frak p\psi)(x)=-i\theta\partial_x\psi_{\lvert x},\quad \psi\in L^2(\R),\; x\in\R.
\label{eq:62}
\end{equation}
Let $\ket{n}$ denote the eigenfunctions of the Hamiltonian $\frak h = \frac 12 (\frak q^2 + \frak p^2)$ of the
quantum harmonic oscillator. They form an orthonormal basis of $L^2(\R)$
and span an invariant dense domain $\dd_S$ of analytic
vectors for the operators $\frak q, \frak p$.
Let $W$ denote the unitary operator from $L^2(\R^2)$ to $L^2(\R)\otimes L^2(\R)$
 defined as
\begin{equation}
Wh_{mn}=\ket{m}\otimes \ket{n} \qquad m,n\in\N
\end{equation}
where $\left\{h_{mn}, m,n\in\N \right\}$ is the basis of $L^2(\R^2)$ spanned by
  Wigner transition functions (see e.g. \cite{Bondia:1988nr} for the explicit
  form of the $h_{mn}$'s). 
One has 
\begin{equation}\label{W1}
 W\ll(x_1)W^*= \frak q\otimes\ii,\quad  W\ll(x_2)W^*= \frak p\otimes\ii .
\end{equation}
As a consequence, for $f\in S(\R^2)$,
\begin{equation}
\label{wpis}
 W\ll(f)W^*=\pi_S(f)\otimes\ii
\end{equation}
where $\pi_S$ is the Schr\"odinger representation
(or the Weyl prescription), namely
\begin{equation}
\label{repscro}
 \pi_S(f)\doteq \int \hat{f}(k_1,k_2)e^{\frac{i}{\theta}(\frak
   q k_1+\frak p k_2)}dk_1 dk_2.
\end{equation}
\begin{prop}
\label{propschro}
$\overline{\pi_S(S(\R^2))}\simeq {\cal K}(L^2(\R))$ is an irreducible
representation of $\,\kk$. 
\end{prop}
Here ${\cal K}(\hh)$ denotes the set of compact
operators on a Hilbert space $\hh$. This last proposition is a standard result, whose prove is recalled for instance in
\cite{Martinetti:2011fko}. 

\subsection{Quantum coordinates}

To summarize, the abstract quantum coordinates operators $q_1$, $q_2$, once
viewed as
operators affiliated to the algebra of compact operators $\kk$, have
two two natural representations: 
\begin{description}
\item[-] a reducible one $\ll(x_1), \ll(x_2)$ on $L^2(\R^2)$,

  \item[-]  an irreducible one, the Schr\"odinger representation, $\pi_S(x_1)
  = \frak q, \;\pi_S(x_2) = \frak p$ on $L^2(\R)$.
\end{description}
\noindent In complex coordinates, 
\begin{equation}
  \label{eq:33}
  z= \frac{x_1 + ix_2}{\sqrt 2},\; \quad\quad \bar z= \frac{x_1 + ix_2}{\sqrt 2},
\end{equation}
one similarly has
\begin{equation}
W\ll(\bar{z})W^*= \mathfrak a^*\otimes\ii,  \quad W\ll(z)W^*= \mathfrak a\otimes\ii\\\label{eq:31bbis}
\end{equation}
where \begin{equation}\label{eq:57}
\mathfrak a\doteq
\frac 1{\sqrt 2}(\frak q + i\frak p),\quad\quad \mathfrak a^*\doteq
\frac 1{\sqrt 2}(\frak q - i\frak p),
\end{equation}
are the creation/annihilation operators, satisfying 
    \begin{equation}
[\mathfrak a, \mathfrak a^*] = \lambda_P^2 \ii.
\label{eq:72}
\end{equation}
So the abstract complex quantum coordinates
\begin{equation}
a \doteq \frac 1{\sqrt 2}
(q_1 + iq_2),\quad  a^* = \frac 1{\sqrt 2} (q_1 - iq_2)
\label{eq:90}
\end{equation}
have representation 
$\ll(z)$, $\ll(\bar z)$ on $L^2(\R^2)$ and $\pi_S(z) = \frak a$, $\pi_S(\bar z) = \frak a^*$ on $L^2(\R)$.

\section{Quantum points}
\label{subsecquantumpoints}

Having shown that the algebra of compact operators $\kk=
\overline{\left(S(\R), \star)\right)}$ plays for a quantum
space the role of the algebra $C_0(\M)$ for a Riemannian manifold, and
remembering that a point $x\in\M$ is nothing but the pure state $\delta_x\in\pp(C_0(\M))$, we take as
``quantum points'' the pure states of 
$\kk$.

One immediately gets that a usual point is not a quantum point. Indeed in
the noncommutative case $\delta_x$ is not longer a pure state, not
even a state, for
\begin{equation}
  \label{eq:60}
  \delta_x(f^*\star f) = (f^*\star f)(x)
\end{equation}
has no reason to be positive (unlike the commutative case where $\delta_x(f^*.f) = \bar f(x) f(x) =
\abs{f(x)}^2\in\R^+$). Said differently, by going to the noncommutative
framework one replaces the pointwise product by the non-local Moyal
product. By this we mean that the evaluation of $f\star g$ at a point $x$ involves the
values of $f$ and $g$ not only at $x$, but on all $\R^2$. This
is what forbids
$\delta_x$ to be a positive linear form. In this precise mathematical
sense, ``points
become fuzzy'' in a quantum space. 

To determine ``how much  fuzzy'', one needs to
work out explicitly the pure state space $\pp(\kk)$. 

\subsection{Quantum points as pure states}

Having single out the Moyal algebra, or equivalently its $C^*$-closure
$\kk$, as the relevant algebra to describe the quantum space,  
it becomes very easy to explain our claim in section
\ref{sectionpoints}, that a ``state'' defined as a positive normalized linear form on
an algebra is the generalization of the ``state'' as it appears in
quantum mechanics. The latter usually denotes the state vector (the
ket) $\ket{\psi}$ in some Hilbert space $\hh$, describing a
quantum state of the physical system under studies. The state, in the sense of a positive application, is simply
the corresponding mean value of observables, namely
\begin{equation}
\omega_\psi(a)\doteq \bra{\psi} a\ket{\psi}.\label{eq:56}
\end{equation}
So to
any \emph{state vector} $\psi$ of quantum mechanics corresponds a
\emph{vector state} $\omega_\psi$ in the sense of $C^*$-algebra (linearity and
positivity of $\omega_\psi$ is obvious, normalization comes from
$\norm{\ket{\psi}}=1$, required by the probabilistic interpretation of
  quantum mechanics). To a
  mixed state in quantum mechanics, characterized by a density matrix
  $\rho = \sum_j p_j \ket{j}\bra j$ in $\bb(\hh)$, corresponds the state
  \begin{equation}
\omega_\rho(a) \doteq  \text{Tr}(\rho a)\label{eq:57bis}
\end{equation}
in the sense of $C^*$-algebra. Furthermore, $\omega_\rho$ is non pure
whereas $\omega_\psi$ - viewed as states of the
$C^*$-algebra $\bb(\hh)$ - is pure. Viewed as a state of
a $C^*$-sub algebra of $\bb(\hh)$, $\omega_\psi$  may be pure or not,
depending whether $\bb$ acts irreducibly on $\hh$.

However it is not true that to any state of a
  $C^*$-algebra corresponds a state vector in $\hh$ or a density
  matrix in $\bb(\hh)$. More exactly, given a $C^*$-algebra $\A$ and a state
  $\varphi$, one can always build a representation $\pi_\varphi$ of
  $\A$ on a Hilbert space $\hh_\varphi$ so that
  that there exists a vector $\ket{\varphi}\in\hh_\varphi$ such that
  $\varphi(a) = \bra{\varphi} \pi_\varphi(a) \ket{\varphi}$ (this is
  the GNS construction). But most often one cannot build a unique representation
$\pi$ on some Hilbert space $\hh$ such that any state $\varphi$ comes
either as a vector state or as a density matrix.{\footnote{See for
      instance the commutative case: the pure state $\delta_x$ is not
     a  vector state in the representation of $C_0(\R)$ on $\hh = L^2(\R)$,
      for there is no $\psi\in\hh$ such that $f(x) =
      \scl{\psi}{f\psi}$. Each pure state $\delta_x$ however is indeed a
      vector state in the $1$-dimensional representation $\pi_x(f)
      \doteq  f(x)$, that is
$\delta_x(f) = \scl{1}{f 1}$ where the scalar product on $\hh_x = \C$ is simply the
multiplication of complex numbers. But for $x\neq y$, the pure state
$\delta_y$ is not a vector state in the $\pi_x$ representation: there
does not exist a real number $\psi\in\hh_x$ such that $f(y) = \scl{\psi}{f\psi}
$ for any $f\in C_0(\R)$.}} However this happens to be true for
  $\A = \kk$.

Indeed, it is a classical result of operator algebra that any
non-degenerate representation{\footnote{That is $\pi(a)\psi =
    0 \;\forall a\in \kk \Longrightarrow a=0$.}} of $\kk$ is unitary equivalent to a
multiple of the unique (up to unitary equivalence) irreducible
representation $\pi_S$ of $\kk$ on $\hh_S$ (the index $S$ is for
Schr\"odinger, see below). 
So limiting ourselves to irreducible representations, we can identify
the abstract $C^*$-algebra $\kk$ with the algebra of compact operators
on $\hh_S$, that is $\pi_S(\kk) = {\cal K}(\hh_S)$ (see
\cite[IV.1.2]{blackadar2006}).
Now, any state $\varphi$  in $\ss(\kk)$
extends to a \emph{normal} state on $\bb(\hh_S)$, still denoted
$\varphi$. ``Normal'' means that there exists a trace-class operator
$s_\varphi\in\bb(\hh_S)$ such that 
\begin{equation}
  \label{eq:41}
  \varphi(a) = \text{Tr} (s_\varphi \, a) \quad \forall a\in\bb(\hh_S).
\end{equation}
If $s_\varphi$ has rank one, $\text{rank}(s_\varphi) =\left\{\ket{\psi}\in\hh_S\right\}$, then
\begin{equation}
\varphi(a) = \text{Tr}(s_\varphi a) = \scl{\psi}{a\psi}
\label{eq:58bis}
\end{equation}
so that $\varphi$ is pure. We summarize these remarks in the following
proposition, using as well proposition \ref{propschro}.
\begin{prop}
Any pure state $\omega$ of the algebra $\kk$ of quantum space-time is a
vector state in the irreducible representation $\pi_S$ on $\hh_S$,
that is there exists a state vector $\psi\in L^2(\R)$ such that
\begin{equation}
  \label{eq:54}
  \omega(f) = \scl{\psi}{\pi_S(f)\psi} \quad \forall f\in S(\R^2).
\end{equation}
\end{prop}

\subsection{Generalized coherent states}

Among the vector states (\ref{eq:54}), we will pay attention to two
particular classes. The first one are those states given by a vector
$\psi\in L^2(\R)$
whose decomposition on the basis $\left\{ \ket{n},\, n\in\N\right\}$
has only one non-zero component, namely the eigenstates of the quantum
harmonic oscillator
\begin{equation}
  \label{eq:64}
  \omega_m(f) \doteq \bra{m}\pi_S(f)\ket{m}.
\end{equation}

Another class of interesting states are the coherent states. Recall \cite{Cohen-Tannoudji:1973fk}
that a coherent (or semi-classical) state of the quantum harmonic
oscillator is a quantum state that reproduces the
behaviour of a classical  harmonic oscillator. The movement of such an
oscillator with given mass $m$ and angular velocity $\omega$ is fully
characterised by one complex number $\kappa  =\abs{\kappa}e^{i\Xi}$ giving the
amplitude of oscillation $\abs{\kappa}$ and the phase $\Xi$. The same is true
for a quantum coherent state (see e.g. \cite{Martinetti:2011fko}).
\begin{defi}
A coherent state of the Moyal algebra $\A$ is a linear form 
\begin{equation}
  \label{eq:3}
 \omega_\kappa^{\text{c}}( f) \doteq  \scl{\kappa}{\pi_S(f)\kappa} \quad
  \forall f\in\A
\end{equation}
where $\ket{\kappa}\in L^2(\R)$, $\norm{\kappa}_{L^2(\R)} =1$, is a solution of 
\begin{equation}
  \label{eq:4}
\frak a \ket{\kappa} = \lambda_P\kappa\ket{\kappa} \quad \kappa\in\C.
\end{equation}
\end{defi}
The development of a coherent states on the basis of eigenstates is 
  \begin{equation}
    \label{eq:7bis}
 \ket{\kappa} = \sum_{m\in\N} c^\kappa_m \varphi_m,\quad c^\kappa_m=
e^{-\frac{\abs{\kappa}^2}2}\frac{\kappa^m}{\sqrt{m!}}.
  \end{equation}
The exists another characterization of a coherent state, in terms of
translation. 

 Given
$\kappa\in\R^2\simeq \C$, we denote $\alpha_\kappa f$ the
translated of $f\in S(\R^2)$, that is  $(\alpha_\kappa f)(z) =
f(z+\kappa)$, and $\alpha_\kappa\varphi$ the $\kappa$-translated of a
state $\varphi$, that is.
\begin{equation}
  \label{eq:65}
  (\alpha_\kappa \varphi)(f) \doteq \varphi(\alpha_\kappa f).
\end{equation}
\begin{prop}
\label{propcoherent}\textnormal{[see e.g. \cite{Martinetti:2011fko}]}
The coherent state $\omega_\kappa^c$ is the translated of the ground
state of the
quantum harmonic oscillator, with translation
$\sqrt{2}\lambda_P\kappa$. That is to say
\begin{equation}
  \omega^c_{\kappa}(f) =  \alpha_{\sqrt{2}\lambda_P\kappa}\omega_0(f).
\label{eq:74}
\end{equation}
\end{prop}
 The coherent states 
are particularly
important for the DFR model since they are the states of optimal
localization, that is those which minimize the uncertainty
\eqref{eq:1} in the
measurement of the coordinates \cite{Doplicher:1995hc}.

In the following, we will consider a larger classes of states.
\begin{defi}
\label{gencos}
We call a \emph{generalized coherent state} any element in $\pp(\kk)$ obtained
by translation of an eigenstate of the Hamiltonian of the quantum harmonic
oscillator. The set of all generalized coherent states is
\begin{equation}
  \label{eq:181}
  {\cal{C}}\doteq \underset{m\in\N}{{\Large{\cup}}}\ccc(\omega_m)\quad \text{ where }\quad
\ccc(\omega_m)\doteq \left\{ \alpha_\kappa\,\omega_m,\; 
    \kappa\in\R^2 \right\}.
\end{equation}
\end{defi}
\section{Quantum length and spectral distance in the Moyal plane}

In this section, we list the results on the quantum length and the
spectral distance between gene\-ralized  coherent states obtained in
\cite{Martinetti:2011fk} for the former, in \cite{Cagnache:2009oe} and
\cite{Martinetti:2011fko} for the latter.

\subsection{Quantum length}
\label{sectionqlength}

It is easier to compute the square root of a mean value, that
the mean value of a square root. So, rather than the quantum length
$d_L$ introduced in \eqref{eq:55} we will consider the
square-root of the quantum square-length
\begin{equation}
  \label{eq:66}
  d_{L^2}(\omega, \tilde\omega) \doteq (\omega\otimes \tilde\omega)(L^2).
\end{equation}
Thanks to the similarities between the length operator and the
Hamiltonian of the quantum harmonic oscillator, $d_{L^2}$ is not difficult to calculate.

\begin{prop} \textnormal{\cite{Martinetti:2011fk,
  Bahns:2010fk}}
\label{propquantdist} The quantum square-length on the set $\ccc$  of
generalized coherent states introduced in definition
\ref{gencos} is
\begin{equation}
  \label{eq:bbb162}
  d_{L^2}(\alpha_{\kappa} \omega_m, \alpha_{\tilde\kappa} \omega_n) =
  2E_m + 2E_n + \abs{\kappa-\tilde\kappa}^2
\end{equation}
for any $m,n\in \N,\; \kappa, \tilde\kappa\in\R^2$, with
\begin{equation}
E_m = \lambda_P^2(m+\frac 12)\label{eq:162}.
\end{equation}
 the $n^\text{th}$ eigenvalue of the
 Hamiltonian $H$ of the quantum harmonic oscillator.
Hence the quantum square length is invariant by translation. Moreover one has \begin{equation}
\label{eq:840}
  d_L(\alpha_\kappa\omega_m,\alpha_{\tilde\kappa}\omega_n) \leq  \sqrt{d_{L^2}(\alpha_\kappa\omega_m,\alpha_{\tilde\kappa}\omega_n)} 
\end{equation}
with equality only when $m=n=0$ and $\kappa=\tilde\kappa$, that is
\begin{equation}
  \label{eq:84}
  d_L(\alpha_\kappa\omega_0,\alpha_\kappa\omega_0) =2\sqrt{E_0} = \sqrt{d_{L^2}(\alpha_\kappa\omega_0,\alpha_\kappa\omega_0)}.
\end{equation}
\end{prop}

\subsection{Spectral distance in the Moyal plane}
\label{section-length}

The known results on the spectral distance in the Moyal plane
are summarized
in the following proposition.
\begin{prop}
\label{propdist}
1.\cite{ Martinetti:2011fko}  The spectral distance
between any state $\varphi\in\sa$ of the Moyal algebra and any of its
$\kappa$-translated, $\kappa\in\C$, is precisely the amplitude of translation
 $$d_D(\varphi, \alpha_\kappa\varphi) = \abs{\kappa}.
 $$
2. \cite{Cagnache:2009oe, Martinetti:2011fko} The spectral distance on the Moyal plane takes all possible
 value in $[0,\infty]$.
\newline

\noindent 3. \cite{Cagnache:2009oe} The distance between eigenstates $\omega_m$ is additive: 
\begin{align*}
d_D(\omega_m,\omega_n)=\frac{\lambda_P}{\sqrt{2}}\sum_{k=m+1}^n{{1}\over{{\sqrt{k}}}}.
\end{align*}
\end{prop}

We stress that there is no misprint at at point 2:  the closing
right bracket indicates that there exist states at infinite distance
from one another. A first example of such states have been exhibited in \cite{Cagnache:2009oe};  other classes of
such states have been worked out in
\cite[Prop. 7]{Cagnache:2009vn}. This is a crucial difference with the $\lambda_P=0$ commutative limit:
both the Moyal plane and $\R^2$ have infinite diameter (i.e. one can
find points/pure states at arbitrarily large distance from
one another), but on $\R^2$ any two points are at finite distance from
one another. 

This also has interesting consequence on the topology of the
state space: the latter is not connected for the metric topology,
while it is connected in the weak* topo\-logy
(that coincides with the topology induced by the trace-norm, see
\cite{Cagnache:2009oe}). In other terms, the topology induced
by the spectral distance is not the weak* topology, meaning that the (minimal unitization) of the Moyal plane is
not a (compact) \emph{quantum metric space} in the sense of Rieffel
\cite{rieffel2003}.

\section{Minimal length and spectral doubling}
Propositions \ref{propquantdist} and \ref{propdist} stress the obvious discrepancy between the quantum length and the
spectral distance:
\begin{align*}
d_D(\omega_m,\omega_n)=\frac{\lambda_P}{\sqrt 2}\sum_{k=m+1}^n{{1}\over{{\sqrt{k}}}}
\; &\quad\neq \quad\sqrt{d_{L^2}(\omega_m,\omega_n)} =\sqrt{2E_m + 2E_n},\\
d_D(\omega_m, \alpha_\kappa\omega_m) = \abs{\kappa} &\quad
\neq \quad \sqrt{d_{L^2}(\omega_m, \alpha_{\kappa} \omega_m)}
= \sqrt{4E_m + \abs{\kappa}^2}.
\end{align*}
So it seems at first sight that the two quantities do not capture the same metric
information on a quantum space, and that it makes little sense to
compare them. There is indeed a fundamental difference: the quantum
square-length $d_{L^2}$ is not a distance, for it does not vanish on
the diagonal (i.e. when the two arguments are equal); unlike the
spectral distance which is a true distance function in the
mathematical sense. We show below how to solve this obvious
discrepancy by
\begin{center}
\emph{turning the quantum length into a true distance}, 

or by

\emph{giving a ``quantum mechanics'' flavor to the spectral
distance}.
\end{center} The two point of view turn out to be equivalent thanks to
the occurrence of a Pythagoras theorem in noncommutative geometry. 
The result is obtained thanks to a standard procedure in
noncommutative geometry, consisting in doubling the spectral triple.

\subsection{Spectral doubling}
Doubling a spectral triple $(\A, \hh, D)$ consists in taking its
product with the
standard spectral triple on $\C^2$ in order to obtain the new
spectral triple
\begin{equation*}
\label{eq:9}
\A'\,\doteq  \A\otimes\, \C^2,\quad\hh'\doteq \hh\,\otimes\,\C^2,\quad
D'\doteq D\,\otimes\,\ii +
\Gamma\otimes D_I
\end{equation*}
where $\Gamma$ is a grading of $\hh$ and 
\begin{equation}
D_I\doteq  \left(\begin{array}{cc} 0&\bar\Lambda\\
\Lambda&0 \end{array}\right)\;\text{ with } \;\Lambda =
\text{ const}.\label{eq:78}
\end{equation}
Pure states of $\A'$ are pairs $\omega^i \doteq (\omega,\delta_i)$
where $\omega$ is a pure state of $\A$ and $\delta_{i=1,2}$ are the pure
states of $\C^2$
\begin{equation}
  \label{eq:73}
  \delta_1 (z_1, z_2)= z_1, \quad \delta_2 (z_1, z_2)=z_2 \quad
  \forall (z_1, z_2)\in \C^2.
\end{equation}
 Hence
 \begin{equation}
\pp(\A') \simeq \pa \times \pa
\label{eq:79}
\end{equation}
and the geometry described by the doubled spectral triple  $(\A',
\hh', D')$   is a \emph{two-sheet model}, with
associated distance $d_{D'}$.

The projection of $d_{D'}$ on each
sheet gives back the distance $d_D$ on a single sheet,
\begin{equation}
d_{D'}(\omega^i, \,\tilde\omega^i) = d_D(\omega, \tilde\omega),
\label{eq:80}
\end{equation}
while the distance between the sheets is constant and non-zero \cite{Martinetti:2002ij,DAndrea:2012fk},
\begin{equation}
d_{D'}(\omega^i, \,\omega^j) = d_{D_I}(\delta^i,\delta^j) = \frac
1{\abs{\Lambda}}.
\label{eq:81}
\end{equation}

The idea is to use this constant $\abs{\Lambda}^{-1}$ to implement the notion of minimal length within the
  spectral distance framework. Namely, rather than comparing the quantum length
  with the spectral distance on a single sheet, one postulates that
  the quantum square-length has to 
be compared with the spectral distance in the double-sheeted model of
quantum space-time. In other terms one aims at identifying
  \begin{equation}
d_{L^2}(\omega, \tilde \omega)\quad \text{ with }\quad d^2_{D'} (\omega^1,
\tilde \omega^2).
\label{eq:82}
\end{equation}
To do so, the free parameter $\Lambda$ is fixed as 
$$\abs{\Lambda}^{-2} = d_{L^2}(\omega, \omega)$$
for some reference state $\omega$, so that 
$d^2_{D'}(\omega^1, \omega^2) = d_{L^2}(\omega, \omega).
$
 The point is then to check whether the identification
 \begin{equation}
{d^2_{D'} \longleftrightarrow d_{L^2}}\label{eq:75}
\end{equation}
holds true for other states. Obviously this has chance to be true only for those
states $\tilde\omega$ belonging to 
\begin{equation}
  \label{eq:76}
 \pp(\omega) \doteq \left\{ \omega \in\pp(\A),\; d_{L^2}(\tilde\omega, \tilde\omega) = d_{L^2}(\omega, \omega)\right\}.
\end{equation}
Luckily,  by proposition \ref{propquantdist} one has that the
translated of any eigenstates $\omega_m$
satisfy the required conditions 
\begin{equation}
  \label{eq:77}
  {\mathcal C}(\omega_m)\subset \pp(\omega_m).
\end{equation}
Furthermore, the distance $d_{D'}$ between two states $\omega,
\tilde\omega\in{\cal C}(\omega)$ localized on different sheets is
known, and given by Pythagoras theorem.

\begin{prop}\cite{Martinetti:2011fko}
The product of the Moyal plane by $\C^2$ is orthogonal in the sense of
Pythagoras theorem, restricted to a set of generalized coherent states
${\mathcal C}(\omega)=\left\{\alpha_\kappa\omega, \kappa\in\C\right\}$:
\begin{equation}
  \label{eq:10bis}
  d_{D'}^2(\omega^1, \tilde\omega^2) =
  d_D^2(\omega, \tilde\omega) +
  d_{D_I}^2(\delta^1, \delta^2)
\end{equation}
for any $\omega, \tilde\omega\in \mathcal{C}(\omega).$
\end{prop}
Therefore, identifying in the double Moyal space
$d^2_{D'}(\omega_m^1, \tilde\omega_n^2 )$ with 
$d_{L^2}(\omega_m, \tilde\omega_n)$ 
amounts to
identifying on a single sheet $d_{D}(\omega_m, \tilde\omega_n)$ with
\begin{equation}
\label{identific}
d'_{L}(\omega_m, \tilde\omega_n)
=   \sqrt{d_{L^2}(\omega_m, \tilde\omega_n) -\abs{\Lambda}^{-2}}.
\end{equation}

\begin{rem}The generalization of Pythagoras theorem to the product of arbitrary
spectral triples has been investigated in \cite{DAndrea:2012fk}. One
finds that for the product of arbitrary unital spectral triples
$(\A_1, \hh_1, D_1)$, $(\A_2, \hh_2, D_2)$, the
following Pythagoras inequalities hold:
\begin{equation}
  \label{eq:12}
d_D^2(\varphi_1, \tilde\varphi_1) + d_{D_I}^2(\varphi_2,
  \tilde\varphi_2)\leq 
   d_{D'}^2(\varphi_1\otimes\varphi_2,\tilde\varphi_1\otimes\tilde\varphi_2)
   \leq 2\lp d_D^2(\varphi_1, \tilde\varphi_1) + d_{D_I}^2(\varphi_2, \tilde\varphi_2)\rp.
\end{equation}
for any $\varphi_1, \tilde\varphi_1\in \ss(\A_1), \varphi_2,
\tilde\varphi_2\in \ss(\A_2)$.
\end{rem}

\subsection{Modified quantum length}

In the analysis above, the free parameter $\Lambda$ has been fixed
once for all by the choice of the reference state $\omega$. If one had started with a reference state
$\omega_0\notin \pp(\omega)$, one would have obtained a result
similar as eq. (\ref{identific}) for any state $\tilde\omega_0\in {\cal C}(\omega_0)$. In order to
collect the results for all possible choices of reference states into
a single formula, it is convenient to introduce the \emph{modified quantum length}
\begin{equation}
d'_L(\omega, \tilde\omega) \doteq \sqrt{\abs{d_{L^2}(\omega, \tilde\omega)
    - \Lambda^{-2}(\omega, \tilde\omega)}}
\label{eq:83}
\end{equation}
where
\begin{equation}
\Lambda^{-2}(\omega, \tilde\omega) = \sqrt{d_{L^2}(\omega, \omega)
  d_{L^2}(\tilde\omega, \tilde\omega)}.\label{eq:85}
\end{equation}
In case $\tilde\omega\in\pp(\omega)$, eq.~(\ref{eq:83}) gives back (\ref{identific}).

\begin{figure}[h*t]
\begin{center}
\vspace{-4.75truecm}\mbox{\rotatebox{0}{\scalebox{.8}{\hspace{3truecm}
\includegraphics{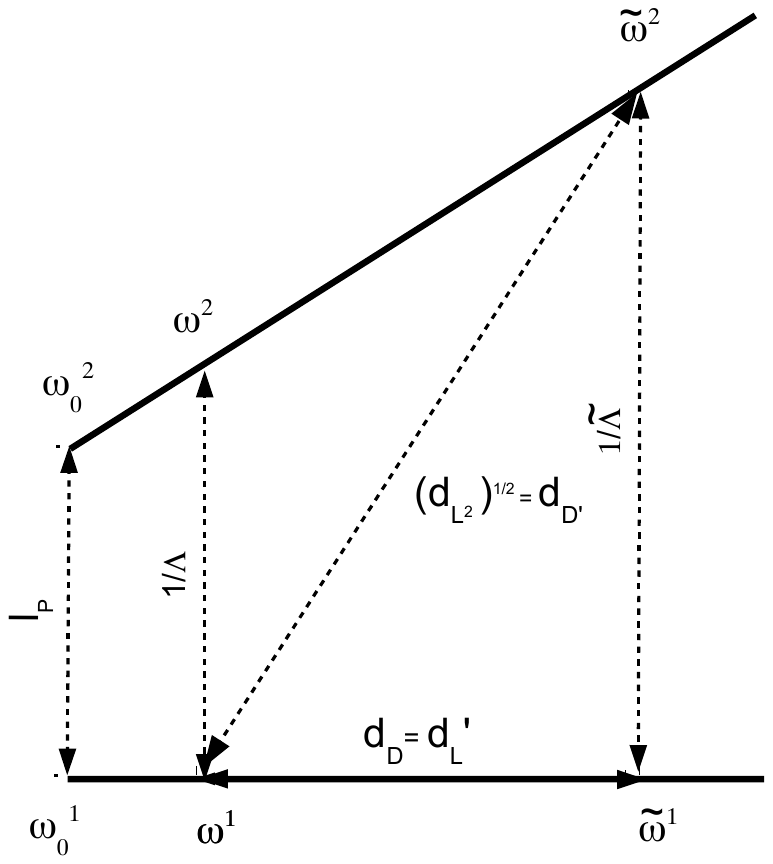}}}}
\end{center}
\vspace{-5.75truecm}\caption{\label{figurecorfou}} The spectral
distance in the double Moyal space and the (modified) quantum length:
here $\omega$ is in $\pp(\omega_m)$ and $\tilde\omega$ in
$\pp(\omega_n)$ for non-zero distinct integers $m,n$.
\end{figure}

The modified quantum length is the correct quantity, build from the
length operator $L$, that should be compared with the spectral distance.
\begin{prop}\cite{Martinetti:2011fk}
\label{propmodifiedql}
On a set of generalized coherent states, ${\cal C}(\omega_m) =
\left\{\alpha_\kappa\omega_m, \kappa\in \C\right\}$, 
for instance the states of optimal localization ${\cal C}(\omega_0)$,
the  identification between the spectral distance and the quantum
length  holds true both in the two-sheet model,
\begin{equation}
\label{theone0}
d_{L^2}(\omega,\tilde\omega) = d^2_{D'}(\omega^1, \tilde\omega^2)
\quad\quad \forall \omega, \tilde\omega\in {\cal C}(\omega_m),
\end{equation}
and on a single sheet (see figure \ref{figurecorfou})
\begin{equation}
d_D(\omega,  \tilde\omega)  = d'_L(\omega,  \tilde\omega) \quad\quad\quad
\forall \omega, \tilde\omega\in {\cal C}(\omega_m).
\label{eq:69}
\end{equation} 

On the set of all generalized coherent states, $d_D$ coincides with
$d'_L$ asymptotically, both in the limit of large translation
\begin{equation}
  \label{eq:185quinte}
  \lim_{\kappa\to \infty} \frac{d_D(\akom, \akton) -d'_L(\akom,
    \akton)}{d'_L(\akom,\akton)} = 0, \quad \forall m,n\in\N,\, \tilde\kappa\in\C,
\end{equation}
 and for large difference of energy
\begin{equation}
  \label{eq:185}
  \lim_{n\to 0} \frac{d_D(\akom, \akton) -d'_L(\akom,
    \akton)}{d'_L(\akom,\akton)} = 0,\quad \forall m\in\N,\, \kappa,\tilde\kappa\in\C.
\end{equation}
\end{prop}

Eq. (\ref{eq:185}, \ref{eq:185quinte}) are analogous to (\ref{eq:69}),
in that they indicate a relation between two quantities on a single
copy of the Moyal plane. There is no analogous
to (\ref{theone0}), namely the identification of the spectral distance
with the modified quantum length has no equivalent in the double Moyal
space. This is because for $m\neq n$, one has $\omega_m\notin
\pp(\omega_n)$ so that to any two sets of generalized
coherent states $\ccc(\omega_m), \ccc(\omega_n)$ correspond two Dirac operators with distinct free
parameters
\begin{equation}
\Lambda= d_{L^2}^{-\frac 12}(\omega_m, \omega_m) \quad\text{  or  }\quad \tilde\Lambda=
d_{L^2}^{-\frac 12}(\omega_n, \omega_n).
\label{eq:86}
\end{equation}
In the almost-commutative
case, that is the doubling of the standard spectral triple associated
to a manifold, this various Dirac operators are collected into a
single operator with non-constant free parameter $\Lambda(x)$. The
latter is interpreted as a Higgs field \cite{Connes:1996fu,Martinetti:2002ij}. At the moment it is not clear whether a
similar procedure can be performed in the Moyal plane.

\subsection{Turning the quantum length into a true distance}

The spectral doubling, consisting in viewing a pair of states $(\tilde\omega,
\omega)$ in $\pp(\kk)$ as living on two distinct sheets, gives a sense to the notion of minimal
spectral distance. From this perspective, it furnishes the ``quantum
taste'' to the spectral distance that we mention at the
beginning of this section. One could also start from
the other assumption,  namely ``turning the quantum length into a true
distance''. The natural way to do so is to make the minimum of the spectrum of the
length operator zero, by defining a modified length operator
\begin{equation}
  \label{eq:38}
  L' \doteq L - l_p \ii\otimes\ii
\end{equation}
with $l_P = d_{L^2}(\omega_0, \omega_0)$ is defined in (\ref{eq:40}). But this guarantees that
\begin{equation}
d_{L'}(\omega,\omega) \doteq (\omega\otimes\omega)(L')= 0
\label{eq:46}
\end{equation}
only for $\omega\in\pp(\omega_0)$. In order to make (\ref{eq:46}) true for all
$\omega\in \pp(\kk)$, one is led quite naturally to the definition
\eqref{eq:83} of the modified quantum length $d'_L$. 

However there is no selfadjoint operator
${L'}^2$ such that
\begin{equation}
{d'_L}^2(\tilde\omega,\omega)\quad \text{ would equal } \quad  (\tilde\omega\otimes\omega)({L'}^2).\label{eq:70}
\end{equation}
For non pure states $\varphi, \tilde\varphi$ this is obvious since
$(\tilde\varphi\otimes\varphi)({L'}^2)$ is linear in $\varphi$ and
$\tilde\varphi$ whereas ${d'_L}^2(\tilde\varphi, \varphi)$ is not. For
pure states, one can check that ${L'}^2$ does not exist by writing a condition, relying on the
linearity of $(\omega\otimes\tilde\omega)({L'}^2)$, that cannot be
satisfied by our definition \eqref{eq:83} of $d'_L$. Let us do it
explicitly. For any  $\;
i,j\in\N$, we write $\omega_{ij}\doteq \omega_{\frac 1{\sqrt
    2}(\ket{i}+ \ket{j})}$ and $\ket{ij} \doteq \ket{i}\otimes
\ket{j}$,  $\; i,j\in\N$. By easy computations one gets
\begin{equation}
  \label{eq:71}
  (\omega_{ij}\otimes\omega_l)({L'}^2) = \re\,\bra{jl} {L'}^2
     \ket{il} + \frac 12 (\omega_i\otimes\omega_l)({L'}^2) + \frac 12
 (\omega_j\otimes\omega_l)({L'}^2)
\end{equation}
so that, for  $\omega_{ijk}\doteq \omega_{\frac 1{\sqrt 3}(\ket{i}+
  \ket{j} + \ket{k})}$,
\begin{equation*}
  \label{eq:37}
  (\omega_{ijk}\otimes\omega_l)({L'}^2) =\frac 13\left( 2\omega_{ij}
    \otimes \omega_l + 2\omega_{ik} \otimes
    \omega_l +
 2\omega_{jk} \otimes \omega_l \right.
    \left.  -\omega_i\otimes \omega_l -
    \omega_j\otimes \omega_l-   \omega_k\otimes \omega_l\right)({L'}^2).
\end{equation*}
Therefore, condition \eqref{eq:70} would imply 
\begin{align}
  \nonumber 3{d'_L}^2(\omega_{ijk},\omega_l) &=2{d'_L}^2(\omega_{ij},\omega_l) +
2{d'_L}^2 (\omega_{ik},\omega_l) +
2{d'_L}^2(\omega_{jk},\omega_l)\\
\label{nonumber}
    &   -{d'_L}^2(\omega_i,\omega_l) -
   {d'_L}^2(\omega_j,\omega_l)-   {d'_L}^2 (\omega_k, \omega_l)
\end{align}
But, noticing that
\begin{equation}
L^2 =2( H\otimes \ii + \ii \otimes H- a\otimes a^* - a^*\otimes a)
\label{eq:100}
\end{equation}
where $a$ has been defined in section~(\ref{eq:90}) and $H = \frac 12
(q_1^2 + q_2^2)$ is the Hamiltonian of the harmonic oscillator, one
gets for $i,j,k,l$ four integers whose differences are greater than
one in absolute value,
\begin{align}
d_{L^2}(\omega_i, \omega_j)= 2E_i + 2E_j, &\quad\quad
d_{L^2}(\omega_{ij},\omega_{kl}) = E_i + E_j + E_k + E_l,\\
d_{L^2}(\omega_{ijk},\omega_{l}) = 2E_l + \frac 23 (E_i + E_j+
  E_k),&\quad\quad
 d_{L^2}(\omega_{ijk},\omega_{ijk}) = \frac 43 (E_i + E_j + E_k)
\label{eq:101}
  \end{align}
with $E_m$ defined in (\ref{eq:162}). Then by \eqref{eq:83}
\begin{align}
  \label{eq:103}
  {d'_{L}}^2(\omega_{i}, \omega_{l}) =  (\sqrt{2E_i}-\sqrt{2E_l})^2, 
&\quad {d'_{L}}^2(\omega_{ij}, \omega_{l}) =  (\sqrt{E_i + E_j}-\sqrt{2E_l})^2,\\
{d'_{L}}^2(\omega_{ijk}, \omega_{l}) &=  (\sqrt{\frac 23 (E_i + E_j +
  E_k)}-\sqrt{2 E_l})^2.
\label{eq:98}
\end{align}
One then easily checks that for this choice of $i,j,k,l$ eq.\eqref{nonumber} does not hold.
 
Consequently there is no modified length operator $L'$ corresponding to
the modified quantum length $d'_L$.
It is quite remarkable that the spectral distance $d_D$ on a single copy of
the Moyal plane coincides (exactly on the set  of translated of a
states, asymptotically on the set of generalized coherent states) with
the ``natural'' quantity $d'_L$, vanishing on
the diagonal, that one can
build from the quantum length $d_L$. The two options
``quantizing the spectral distance'' by allowing the emergence of a
non-zero minimal spectral distance,  or ``geometrizing the quantum
length'' by turning it into a true distance are two equivalent
procedures.

\section{Geodesics in the Moyal plane}

In a quantum space, there is no natural notion of geodesics. However
we may find a substitute in the notion of \emph{optimal element} that
we introduced in section in section \ref{sectiongeodesic}. Recall that given a spectral triple $(\A, \hh, D)$,  by optimal element between two
states we intend an element of the algebra that attains the supremum
in the spectral distance formula, or a sequence of elements in case
the supremum is not attained. Noticing that the commutator norm
condition can be equivalently written as an equality instead of an
inequality \cite{Iochum:2001fv}, an optimal element between
$\tilde\varphi,\varphi\in\sa$ is thus either an element of $\A$ such that 
\begin{equation}
  \label{eq:95}
  \abs{\tilde\varphi(a) - \varphi(a)} = d_D(\tilde\varphi, \varphi) \quad
  \text{ and } \quad \norm{[D,a]}=1,
\end{equation}
or a sequence of element $a_n\in\A$ such that
\begin{equation}
  \label{eq:950}
  \lim_{n\to\infty} \abs{\tilde\varphi(a_n) - \varphi(a_n)} = d_D(\tilde\varphi, \varphi) \quad
  \text{ and } \quad \norm{[D,a_n]}\leq 1 \quad \forall n\in\N.
\end{equation} 
For non-unital spectral triples (e.g. non-compact manifolds), one
usually finds first an element that satisfies
(\ref{eq:95}) but which is not in the algebra and needs to be
``regularize at infinity'', like the
function (\ref{eq:18}) is approximated by the sequence
(\ref{eq:20}). In this case, as explained in section
\ref{sectiongeodesic}, we talk about an optimal element up to
regularization. In the following we are not interested in the
regularization procedure, and optimal element always means ``up to
regularization''. 

In the commutative case, the commutator norm
condition
\begin{equation}
\norm{[\ds,f]} = \suup{x\in\M}{\norm{{\nabla f}_{\lvert
    x}}_{T_x \M}} = 1
\label{eq:58}
 \end{equation}
characterizes the optimal element between $\delta_x$ and $\delta_y$ locally, in the
 sense that the constraint is carried by the gradient of $f$. The geodesics through $x$
are retrieved as the curves tangent to the optimal element
$f=d_{\text{geo}}(x,\, .)$.
In this sense, computing the
spectral distance amounts to solving the equation of the geodesics:
\begin{enumerate}
\item[-] eq. (\ref{eq:58}) plays the role of the geodesic equation;

 \item[-] the optimal element $f=
   d_{\text{geo}}(x,\, .)$ fully characterizes the geodesics through $x$;

\item[-] the valuation of the optimal element on $\delta_x - \delta_y$ gives the integration of
  the line element on a minimal geodesic between $x$ and $y$.
\end{enumerate}
For these reasons, as a proposal for a ``geodesic'' between two quantum
points $\varphi, \tilde\varphi$, we shall draw our attention on the
optimal elements.
 
\subsection{Discrete versus continuous geodesics}

For eigenstates of the quantum harmonic oscillator whose diffe\-rence of energy $E_n-E_m$  ($m\leq n$ to fix
notation) is small, the spectral distance
\begin{equation}
  \label{eq:125}
    d_D(\omega_m,
\omega_n) =\lambda_P \sum_{k=m+1}^n{{1}\over{{\sqrt{2k}}}}
\end{equation}
 appears as a middle Riemann sum
approximation of the modified quantum length
\begin{equation}
  \label{eq:124}
 d_L'(\omega_m,\omega_n) = \sqrt{2E_n} -\sqrt{2E_m} = \lambda_P\left(\sqrt{2n+1} -
  \sqrt{2m+1}\right) = \lambda_P\int_{m+ \frac 12}^{n+\frac 12} \frac 1{\sqrt{2k}}dk.
\end{equation}

From a geometrical point of view, one may interpret this result saying
that the spectral distance and the quantum length are the integration
of the same quantum line
element
\begin{equation}
\lambda_P\frac 1{\sqrt{2k}}dk
\label{eq:87}
\end{equation}
but along two distinct geodesics: a
continuous one for the quantum length, a discrete one for the spectral
distance.  In a word, \emph{both the spectral distance and the length
operator quantize the line element; with the spectral distance one
also quantizes the geodesics}.

Let us develop this idea from the point of view of the optimal
element. Recall that on the Euclidean plane, the
function $l(x_\mu)$ in (\ref{eq:10})
yields both the length operator $L=l(dq_\mu)$  
and the optimal element
$l(q_\mu)$ between any two pure states $\delta_x, \delta_{\lambda x}$, $
\lambda \in \R^+$. This is no longer true in the Moyal
plane. To see it, it is convenient to work with the complex
coordinates introduced in (\ref{eq:33}), as well as with their universal differential
\begin{equation}
  \label{eq:96}
 da = \frac 1{\sqrt 2}(dq_1 + idq_2), \quad  da^* = \frac 1{\sqrt 2}(dq_1 - idq_2).  
\end{equation}
 \begin{prop}\cite{Martinetti:2011fk}{\footnote{Notice some change of notations with respect to
\cite{Martinetti:2011fk}:  there we assumed that $\ll(l_0)$ were
$l_0(a)$, but there is no guaranty that this should be true. Also
we used indistinctly  $l_i$ for $l_i(a)$, which might have been confusing.}}
\label{moyalgeo}
On the Moyal quantum plane, the length operator can be equivalently
defined as  
$ L=l_i (da)$, 
with
\begin{equation}
l_1(z)
\doteq  \sqrt{z \bar z + z\bar z} 
\;\text{ or } \; l_2(z) \doteq  \sqrt{2(z \bar z  - \lambda_P^2)} \;\text{ or } \;l_3(z)
\doteq  \sqrt{2(\bar z z + \lambda_P^2)}. 
\label{eq:158}
\end{equation}
The optimal element between any two eigenstates of the Hamiltonian of the
quantum harmonic oscillator is - up to regularization at infinity - 
$\ll(l_0)$
where $l_0$ is a solution of 
\begin{equation}
  \label{eq:1590}
  \left(\partial_z l_0 \star z\right) \star \left(\partial_z
    l_0 \star z\right)^*= \frac 1{2} z^*\star z.
\end{equation} 
Neither $l_{1}(a)$ nor $l_{2}(a)$ or $l_{3}(a)$ are optimal elements
between eigenstates.
\end{prop}

If $l_{1}(a)$ were the optimal element, then the identification
between the modified quantum length $d'_L$ and the spectral distance $d_D$  on the set eigenstates of
the harmonic oscillator, discussed in proposition \ref{propmodifiedql}, would
hold true exactly and not only asymptotically. Indeed one checks that \cite{Martinetti:2011fk}
\begin{equation}
\abs{\omega_m (l_{1}(a)) -\omega_n (l_{1}(a))} =\lambda_P\abs{\sqrt{2m+1} -
  \sqrt{2n+1}} = d'_L(\omega_m, \omega_n).
\label{eq:129}
\end{equation}
We may interpret this equation as a definition of 
an ``optimal element for the modified
quantum length'', namely we assume that (\ref{eq:129}) is the
supremum of $\omega_m - \omega_n$ on the unit ball of $\kk$ for some
(still to determine)  semi-norm, distinct from $\norm{[D,.]}$. Having
in minds that optimal elements provide a notion of geodesics in a
quantum space, $\ll(l_0)$ and $l_1(a)$ thus appear as
two proposals for a geodesic on the quantum space, with associated geodesic distance
$d_D(\omega_m,\omega_n), d'_L(\omega_m,\omega_n)$.

\subsection{Shift vs. identity}

Let us now consider translated states.
The function 
\begin{equation}
  \label{eq:133}
  l_\kappa(z) = \frac{ze^{-i\Xi} + \bar ze^{i\Xi}}{\sqrt 2}, \quad \text{ with
  } \quad \Xi \doteq \text{Arg} \,\kappa,
\end{equation}
yields the optimal element (up to regularization at infinity) between any state $\varphi$ and its
$\kappa$-translated both on the Euclidean plane (through the pointwise
action of $l_\kappa$) and the Moyal plane (through its $\star$-action). For the
latter, this has been shown in \cite[Theo. III.9]
{Martinetti:2011fko}, for the former in
\cite[Prop. 3.2]{dAndrea:2009xr}). In particular
$l_\kappa(z)$ is an optimal element between $\delta_x$ and  $\delta_y$ viewed
as the $\kappa=\frac{y_1-x_1 +i(y_2-x_2)}{\sqrt 2}$-translated of $\delta_x$:
one the one hand, 
\begin{equation}
  \label{eq:89}
\abs{\delta_x( l_\kappa) - \delta_{x+\kappa}(l_\kappa)}=  l_{\kappa}(\kappa) =  \sqrt 2
 \abs{\kappa} = \abs{x-y},
\end{equation}
on the other hand
\begin{equation}
  \label{eq:143}
  [\ds, l_\kappa ] =-i\sqrt 2 \left(\begin{array}{cc} 0 & \bar\partial
      l_\kappa\\ \partial l_\kappa&0\end{array}\right) = -i\left(\begin{array}{cc} 0 & e^{i\Xi}\\ e^{-i\Xi}&0\end{array}\right)
\end{equation} 
 has obviously norm $1$. 
Notice that
\begin{equation}
  \label{eq:88}
  [\ds,l_\kappa]^*[\ds,l_\kappa ] = \ii.
\end{equation}

It is quite remarkable that the same function $l_\kappa$ gives an optimal element
between translated states, regardless of the commutativity of the
algebra. In a sense, \eqref{eq:143} indicates that in both the
Euclidean and the quantum planes, the derivatives
of the optimal element between translated states is proportional to
the identity, meaning that the ``geodesic'' is smooth. Quantum versus
classical is not relevant.
\newline
 
Let us now re-examine the optimal element $\ll(l_0)$ between eigenstates of the
quantum harmonic oscillator.  Modulo regularization at infinity, it can be characterized \cite[Prop. 3.7]{Cagnache:2009oe} as a solution
of
\begin{equation}
  \label{eq:143bis}
  [\ds,\ll(l_0)] =-i\left(\begin{array}{cc} 0 & S^*\\ S&0\end{array}\right),
\end{equation} 
where $S$ is the shift operator (eq.~(\ref{eq:1590}) actually follows
from it). One has
\begin{equation}
  \label{eq:880}
 \ii -  [\ds, \ll(l_0)]^*[\ds, \ll(l_0)] = e_0
\end{equation}
where $e_0$ is the projection on $h_0$. Eq. (\ref{eq:143bis}) indicates that the derivative of the
optimal element for the spectral distance between eigenstates is the shift $S$, meaning that the
``geodesic'' is discrete. 

Notice that  what prevents $\ll(l_0)$ to satisfy
(\ref{eq:88}) is that the set of eigenstates of the harmonic
oscillator - identified to $\N$ - is not a group (unlike
the set of translated states). The shift acting on $l^2(\N)$ is not a
unitary operator, so that the
optimal element between eigenstates verifies \eqref{eq:880} instead of
\eqref{eq:88}. The latter would be verified if one could
take into account states with negative energy (the shift on $l^2(\zz)$ is unitary).

\section{Conclusions and outlook}

There is no \emph{quantum} standard meter: 
 the DFR and
$\theta$-Minkowski length operator
$L$ make a minimal length emerge from the Moyal plane, 
on the contrary  Connes's
spectral distance provides the same Moyal plane with a metric structure
that does not imply any minimal distance. 

Because of this discrepancy, stemming from the non-zero minimum $l_P$ of the spectrum of
$L$ opposed to the continuum of value $[0,\infty]$ taken by the spectral
distance $d_D$ on the Moyal plane,  there is no obvious way to compare
these two approaches. However, one can extract from the length operator a quantity
$d'_L$ - the modified quantum length - that
 coincides exactly with the spectral distance $d_D$ on any set
$\ccc(\omega_m)$ of generalized coherent states, and asymptotically on
 their union $\ccc=\underset{m\in\N}{\cup}\ccc(\omega_m)$.

Thanks to Pythagoras theorem for the product of spectral triple, this
way of turning the quantum length $d_L$ into a true distance $d'_L$ is
equivalent to implement a minimal non-zero length into the spectral
distance framework by doubling the spectral triple. 
 
As a tentative physical interpretation, we stress that a pair of states $(\varphi, \tilde \varphi)$ can be viewed either as
two states of a single system, or as one  state
$\varphi\otimes\tilde\varphi$ of a two-point system.
The spectral distance $d_D$ measures the distance between the two
states of the same system, hence $d_D(\varphi, \varphi) = 0$ (no
difference between a system in a state $\varphi$, and the same system
in the same state $\varphi$).
On the contrary, two copies of the same system can be in the same quantum state $\varphi$,
yet, they are two distinct copies. Hence $d_L(\varphi,\varphi) =
(\varphi\otimes\varphi)(L)\neq 0$.
By doubling the spectral triple, one reconciles the two points of
view: a pair of quantum points $(\omega, \tilde\omega)$ in $\pp(\kk)$
can be equivalently
seen as 

\begin{enumerate}
 \item[-] a state $\omega\ot\tilde\omega$ of $\pp(\kk) \otimes \pp(\kk)$, on
  which one evaluates the length operator;

\item[-] a pair of states $\left(\omega^1, \tilde\omega^2\right)$ in
  $\pp(\kk)\otimes\pp(\C^2)$, between which one computes $d_{D'}$.
\end{enumerate}

\noindent For this to make sense,  the correct
objects to compare are either
\begin{enumerate}
\item[-] the double-sheet spectral distance $d_{D'}$ with
   the quantum square-length $d_{L^2}$,
\end{enumerate}

or, equivalently thanks to Pythagoras
theorem 

\begin{enumerate}
\item[-] the single-sheet spectral distance $d_D$ with
  the modified quantum length $d'_L$.
\end{enumerate}

The discrepancy between the corrected quantum length and the spectral
distance that remains between eigenstates of the harmonic oscillator with
a small difference of energy has a natural interpretation in terms of
integrations of the same noncommutative line element along two distinct
geodesics: a discrete geodesic for spectral distance, a continuous one
for the corrected quantum length. . 

As outlook, let us mention the following points: 
\begin{itemize}

\item[-] there is a recent result of Wallet \cite{Wallet:2011uq}
on homothetic transformation of the Moyal plane.
To study the
renormalizability of quantum field theory on noncommutative
spacetimes, Grosse and Wulkenhaar had added an harmonic term $\Omega
x^2$ to the Lagrangian. The spectral distance
computed with the corresponding Dirac operator $D_{\Omega}$ is
\begin{equation*}
  \label{eq:3ter}
  d_{D_\Omega} = \frac 1{\sqrt{1 + \Omega^2}} d_D,
\end{equation*}
where $D$ is the Dirac operator corresponding to the theory without harmonic term. This indicates an intriguing  link between renormalizability and the metric structure of space-time.

\item[-] besides the length operator and the spectral distance, there exists
(at least) a third proposal for a quantized version of the distance in
the physics literature, namely the length operator in loop quantum
gravity \cite{Bianchi:2008uq}. Its definition relies on a crucial way
on the holonomy of a suitable connection (Wilson loops). Interestingly, the holonomy of a connection also appears in the
spectral distance formula when one considers the so-called
\emph{fluctuation of the metric} for a spectral triple based on a
algebra of matrix valued functions $C_0(\M)\otimes M_n(\C)$.  The space of
pure state $\pp$ is a $U(n)$- trivial bundle over $\M$, and the connection
associated to a covariant Dirac operator defines on $\pp$ an
horizontal distribution in the sense of sub-Riemannian geometry. The difference between the spectral
distance and the horizontal distance associated to the connection
heavily depends on the size of the holonomy
group \cite{Martinetti:2006db, Martinetti:2008hl}. So it would be
interesting to compare these two metric interpretations of the
holonomy: loop quantum gra\-vity and noncommutative geometry.

\item[-] the Heisenberg group, which comes out naturally in our context as the exponential of the quantum
coordinates $q_\mu$, can also be seen as a sub-Riemannian geometry \cite{Montgomery:2002yq}. By
defining an appropriate covariant Dirac operator as in the preceding remark, one could provide
the Heisenberg group with a spectral distance. It would be interesting
to understand whether this distance is similar as the one coming from
the spectral triple of the Moyal plane.
\end{itemize}
 
\noindent {\bf Acknowledgments}
Work supported by an ERG-Marie
Curie fellowship 237927 \emph{Noncommutative geometry and quantum
gravity} and by the ERC Advanced Grant 227458 OACFT \emph{Operator Algebras and
Conformal Field Theory}.

\appendix
\section{Appendix}
\subsection{Exact Poincaré covariance vs. deformed Poincaré
  invariance}

In the construction of section \ref{eq:3quat} leading to the Moyal
algebra, there is still some freedom in the choice of the symplectic
form $\sigma$. The latest is constrained by the
transformation law of the commutator relation (\ref{eq:7}) under  Poincaré
transformations. Explicitly, identifying operators with
their irreducible representation so that (\ref{eq:7}) reads
\begin{equation}
  \label{eq:51}
  [q_\mu,q_\nu] = i\lambda_P^2 Q_{\mu\nu} = i\lambda_P^2\, \theta_{\mu\nu}\I,
\end{equation}
 we assume there is a unitary
representation of the Poincaré group $\R^d \ltimes O(d-1,1)$ such that 
\begin{equation}
q_\mu\mapsto q'_\mu \doteq \Lambda_\mu^\alpha q_\alpha + a_\mu
\mathbb{I}\label{eq:6} \quad \quad \Lambda\in
SO(d-1,1), a\in \R^d.
\end{equation}
The commutation relation (\ref{eq:51}) is obviously not
Poincaré invariant since 
 \begin{equation}
  \label{eq:13bis}
  [q'_\mu, q'_\nu] =  
 [a_\mu, a_\nu]\I  +  q_\alpha([a_\mu\I, \Lambda_\nu^\alpha] -
 [a_\nu\I,\Lambda_\mu^\alpha]) + \Lambda_\mu^\alpha[q_\alpha,
 q_\beta]\Lambda_\nu^\beta \neq [q_\mu, q_\nu].
\end{equation}
However Poincaré covariance can be restored observing that the
generators of the classical Poincaré group commute with each other, in which
case (\ref{eq:13bis}) reduces to
 \begin{equation}
  \label{eq:13}
  [q'_\mu, q'_\nu] =  i\lambda_P^2\,\Lambda_\mu^\alpha
  \Lambda_\nu^\beta \theta_{\alpha\beta}\I =  i\lambda_P^2 \left( \text{Ad}_\Lambda \Theta\right)_{\mu\nu}\I. 
\end{equation}
In other term the commutator relations (\ref{eq:51}) are covariant under
Poincaré transformations as soon as one requires the matrix
$\Theta=\left\{\theta_{\mu\nu}\right\}$ to transform under the adjoint action of
the Poincaré group. This requirement is the building block of the DFR
model of \emph{Poincaré covariant} quantum spacetime.

Alternatively, one may
impose the commutators to be invariant under the action of the symmetry
group of the quantum space. This forces to
deform the Poincaré group into the quantum group $\theta$-Poincaré \cite{Amelino-Camelia:2010fk}. The
latest is characterized by a
non-trivial commutation relation between the generators of
translations, 
\begin{equation}
  \label{eq:52}
[a_\mu, a_\nu] = i\theta_{\mu\nu} -
i\theta_{\alpha\beta}\Lambda_\mu^\alpha \Lambda_\nu^\beta,  
\end{equation}
so that (\ref{eq:13bis}) yields
\begin{equation}
  \label{eq:53}
   [q'_\mu, q'_\nu] =  i\lambda_P^2 \theta_{\mu\nu}\I.
\end{equation}
This is a model of \emph{deformed-Poincaré invariant} spacetime,
called  \emph{canonical noncommutative spacetime} (NCS) or 
$\theta$-Minkowski.

 Note that in both DFR and $\theta$-Minkowski the numerical value of the
Planck length is an invariant under symmetry transformations: either because
(\ref{eq:53}) is indeed invariant in $\theta$-Minkowski; or because
in the ($4$ dimensional) DFR model $\lambda_P$ is retrieved as the
norm of the tensor $Q_{\mu\nu}$, 
which is a Poincaré invariant
quantity. 

To some extent, the invariant deformed-Poincaré
NCS can be viewed as the covariant-Poincaré DFR restricted to one point on the
orbit $\Sigma = \left\{\text{Ad}_\Lambda \Theta, \Lambda\in SO(d-1,1)\right\}$. This is discussed at length in
\cite{Piacitelli:2010fk}. From our purposes, once fixed the matrix
$\Theta$ then the relevant
$C^*$-algebra is $\kk$. For the full DFR model one should take into account the
action (\ref{eq:6}) of the Poincaré group. The
  relevant algebra is then (see \cite{Doplicher:1995hc} for
the original argument, \cite{Piacitelli:2010uq} for a recent presentation) 
 \begin{equation}
  \label{eq:14}
  \E = C_0(\Sigma, \kk) = C_0(\Sigma)\otimes \kk,
\end{equation}
namely the $C^*$-algebra of $\kk$-valued smooth functions vanishing at
infinity. 

\subsection{Pair of quantum points}

In the commutative case, a point $x$ of $\R^d$ is a pure state $\delta_x$ of $C_0(\R^d)$. Similarly, we take as a ``quantum point'' a pure
 state of the algebra $\kk$ (in $\theta$-Minkowski ) or $\E$ (in the
 DFR model). Pure states of $\E$ are couples
\begin{equation}
\omega_S\doteq (\delta_S, \omega)\quad \text{ with } \quad \delta_S
\in {\mathcal P}(C_0(\Sigma)) \simeq \Sigma,\; \omega\in {\mathcal P}(\kk).\label{eq:102}
\end{equation}
A pair of quantum points $(\omega_S,\, \tilde\omega_{\tilde S})$
defines a two-``quantum point'' state $\omega_{S}\otimes\tilde\omega_{\tilde S}$. The latter is a pure
state of the tensor product of complex algebras $\ee\otimes\ee$. However, to guarantee that 
 \begin{equation}
[q_\mu\otimes \ii, q_\nu \otimes \ii] =[\ii\otimes q_\mu, \ii
\otimes q_\nu] =i\lambda_P^2 Q_{\mu\nu}(\ii\otimes\ii)
\label{eq:107}
\end{equation}
(that is, the commutators of the coordinates of two independent quantum points are equal), it has been proposed in \cite{Bahns:2003fk} that the tensor product
$\ee\otimes_{C_0(\Sigma)}\ee$ over the center $C_0(\Sigma)$ of $\ee$ should be used instead. This has the following importance consequence:
 \begin{prop}
\label{etatquotient}
\textnormal{[see e.g. \cite{Martinetti:2011fk}]}
Pure states of $\ee\otimes_{C_0(\Sigma)}\ee$ are pairs $(\omega_{S}, \tilde\omega_{S})$ composed of two pure states of $\ee$ corresponding to the same point
$S\in\Sigma$.
\end{prop}
Consequently, for the DFR model, $\theta$-Minkowski and the Moyal
plane, a pair of quantum points is a pair of pure-states $(\omega,
\tilde\omega)$  of $\kk$. So from our length/distance perspective,
these three models of quantum spaces are equivalent.
\newline

Let us
mention that  in loop quantum gravity, the behaviour under Lorentz transformations of the
minimum of a quantum
observable (specifially: the area)  has been
investigated in \cite{Rovelli:2003uq}. It could be interesting to see
whether the analysis developed would make sense in this context. 
\bibliographystyle{abbrv}
\bibliography{/Users/pierremartinetti/physique/articles/Bibdesk/biblio}
\end{document}